\documentclass{elsart}
\usepackage{graphics}
\usepackage{graphicx}
\usepackage{lscape}
\usepackage{amssymb}
\usepackage{latexsym}
\usepackage{verbatim}
\usepackage{multirow}
\usepackage{lscape}
\journal{Chemical Physics}
\begin{document}

\begin{frontmatter}

\title{Electronic and optical properties of families of polycyclic aromatic 
hydrocarbons: a systematic (time-dependent) density functional theory study}

\author[label1]
{G.~Malloci\corauthref{cor}},
\corauth[cor]{Corresponding author. Tel: 
+39\textendash070\textendash675\textendash4839;
Fax: +39\textendash070\textendash510171} 
\ead{giuliano.malloci@dsf.unica.it}
\author[label1,label2]{G.~Cappellini},
\author[label2]{G.~Mulas},
\author[label1]{A.~Mattoni}
\address[label1]{CNR\textendash IOM and Dipartimento di Fisica, 
Universit\`a degli Studi di Cagliari, Cittadella Universitaria, 
Strada Prov.~le Monserrato\textendash Sestu Km~0.700,
I\textendash09042 Monserrato (CA), Italy}
\address[label2]{INAF \textendash{} Osservatorio Astronomico 
di Cagliari\textendash Astrochemistry Group, Strada 54, Localit\`a 
Poggio dei Pini, 
I\textendash09012 Capoterra (CA), Italy}

\begin{abstract}
Homologous classes of Polycyclic Aromatic Hydrocarbons
(PAHs) in their crystalline state are among the most
promising materials for organic opto-electronics. 
Following previous works on oligoacenes we present a systematic comparative 
study of the electronic, optical, and transport properties of oligoacenes, 
phenacenes, circumacenes, and oligorylenes. Using density functional theory 
(DFT) and time-dependent DFT we computed: (i) electron affinities and first 
ionization energies; (ii) quasiparticle correction to the highest occupied 
molecular orbital (HOMO)-lowest unoccupied molecular orbital (LUMO) gap; 
(iii) molecular reorganization energies; (iv) electronic absorption spectra
of neutral and $\pm1$ charged systems. The excitonic effects are estimated 
by comparing the optical gap and the quasiparticle corrected HOMO-LUMO 
energy gap. For each molecular property computed, general trends as a 
function of molecular size and charge state are discussed.
Overall, we find that circumacenes have the best transport properties, 
displaying a steeper decrease of the molecular reorganization energy at 
increasing sizes, while oligorylenes are much more efficient in absorbing 
low\textendash energy photons in comparison to the other classes.

\end{abstract}


\end{frontmatter}
  
\section{Introduction}
\label{introduction}

A large class of conjugated $\pi$\textendash electron systems, polycyclic 
aromatic hydrocarbons (PAHs) are of fundamental importance in many research 
areas of chemistry as well as in materials science and astrophysics. PAHs
are precursors of extented carbon networks \cite{cur93,pop93} and their carbon 
skeletons can be seen as small pieces of graphene \cite{jia08,san09}. Some 
PAHs are of interest in environmental chemistry due to their carcinogenicity 
and their
ubiquity as air pollutants produced by the combustion of organic matter 
\cite{lun05,reh05}. Neutral and charged PAHs up to large sizes are thought to 
be abundant in space and are thus seen as an intermediate stage between the gas 
and dust phases of the matter between stars \cite{tie05,tie08}. 

Homologous classes of PAHs in their crystalline state are among the most 
promising materials for organic electronics \cite{ant08}. Oligoacenes and their 
derivatives, for example, are being increasingly used as active elements in a 
variety of opto\textendash electronic devices such as organic 
thin\textendash film field\textendash effect transistors \cite{hal02}, 
light\textendash emitting diodes \cite{kan04}, photovoltaic cells \cite{yoo04}, 
and liquid crystals \cite{shi03}. Organic electronics based on functionalized 
acenes and heteroacenes is presently a very active field of research 
\cite{ant06,mur07,hua09,kar09,dea10}.

The main features controlling organic semiconducting devices are exciton 
formation, migration, and dissociation, charge transport, charge collection 
at the electrodes, molecular packing in the bulk material, and absorption and 
emission properties \cite{bre04,san07}. Since many of these properties pertain 
mostly to the molecular units, systematic quantum\textendash chemical 
calculations on isolated PAHs can significantly contribute to our understanding 
of the electronic and optical properties of families with promising 
opto\textendash electronic applications. 

As a part of a more extensive research on PAHs \cite{mal07b}, and following our 
previous work on oligoacenes \cite{mal07,cap09}, we present in this paper a 
comprehensive comparative theoretical study of four homologous classes of PAHs, 
namely oligoacenes, phenacenes, circumacenes, and oligorylenes, in their 
neutral, cationic, and anionic charge\textendash states. This choice of 
families was motivated by the availability of reliable experimental data for 
the first members of each class. They include both catacondensed (oligoacenes
and phenacenes) and pericondensed (circumacenes and oligorylenes) species, 
spanning a representative range of regular, relatively symmetric PAHs. Moreover
circumacenes and oligorylenes converge, at the infinite limit, to a zigzag and 
arm\textendash chair edged graphene nanoribbon, respectively, and as such are 
promising candidates for organic and molecular electronics \cite{per09}. We 
restricted ourselves to the first five members of each class. 

\begin{figure}[b!]
\includegraphics[height=0.55cm]{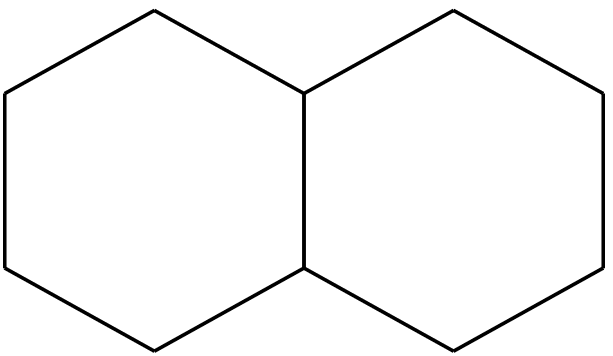}
\includegraphics[height=0.55cm]{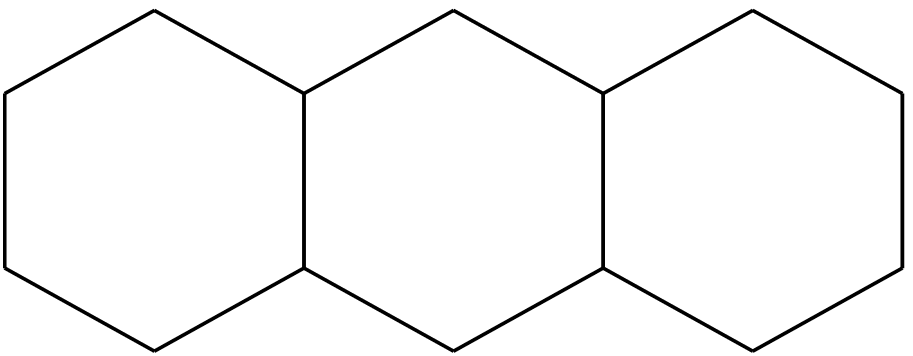}
\includegraphics[height=0.55cm]{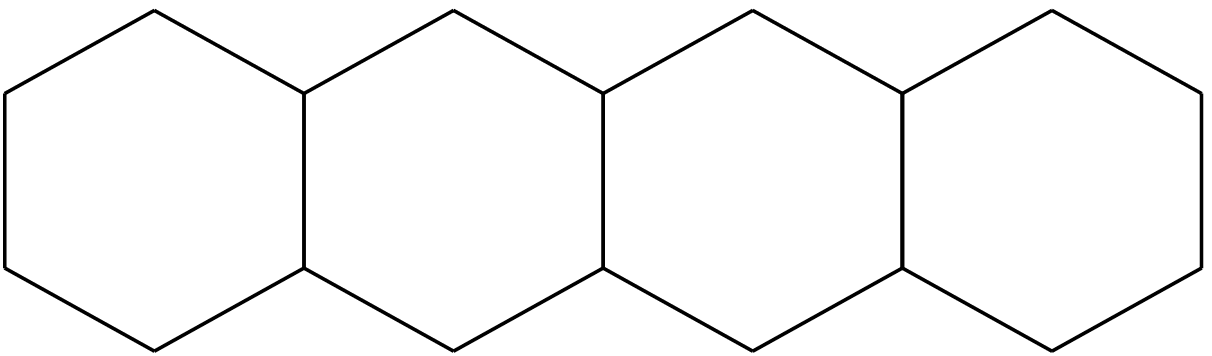}
\includegraphics[height=0.55cm]{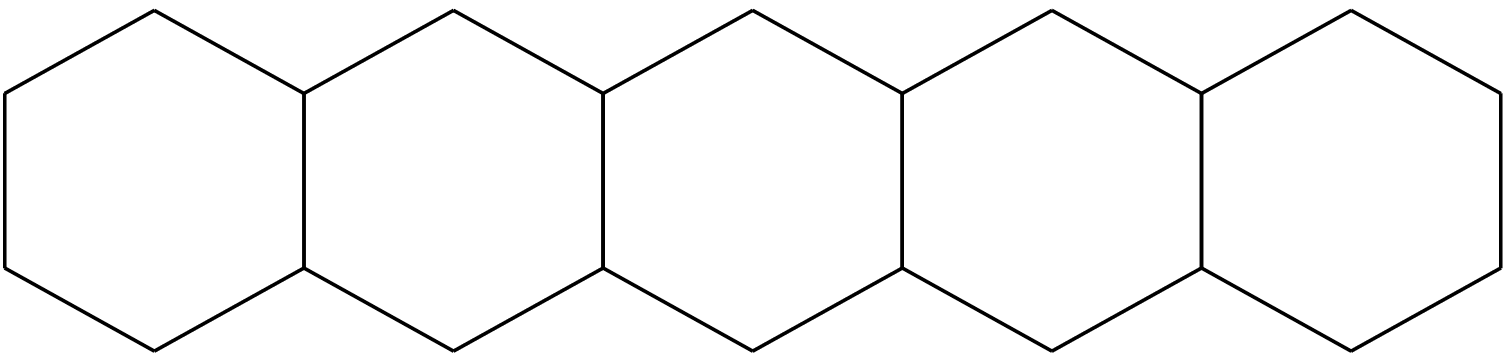}
\includegraphics[height=0.55cm]{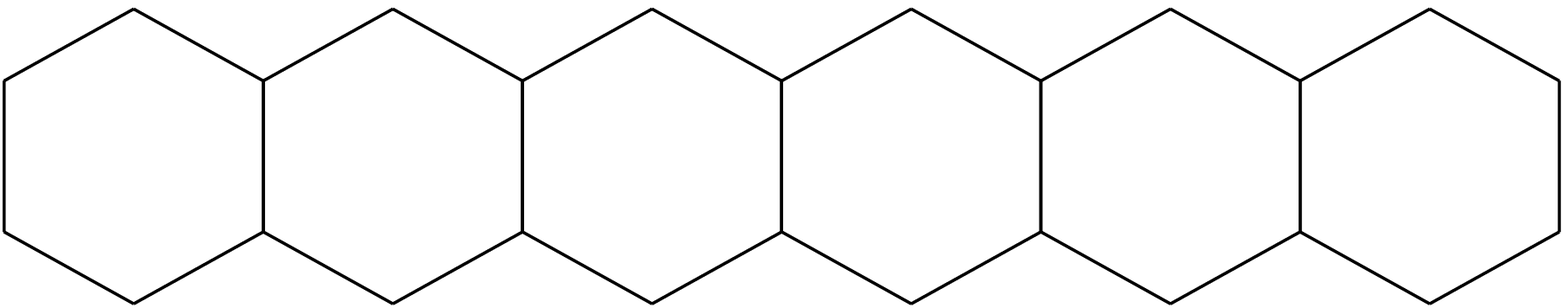}\\[2pt]
\includegraphics[height=0.7cm]{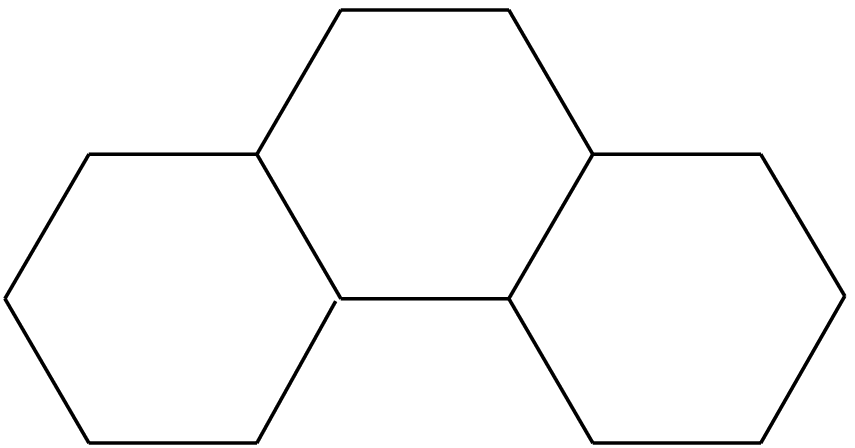}
\includegraphics[height=0.7cm]{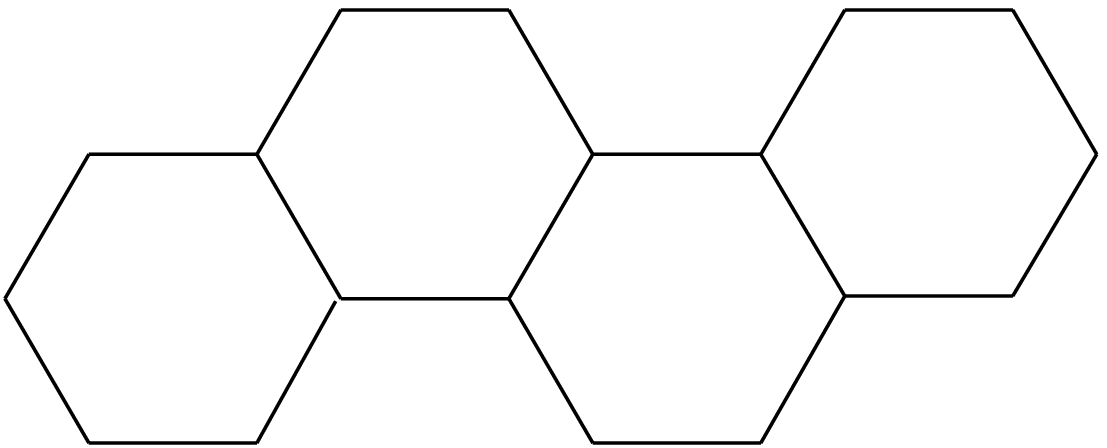}
\includegraphics[height=0.7cm]{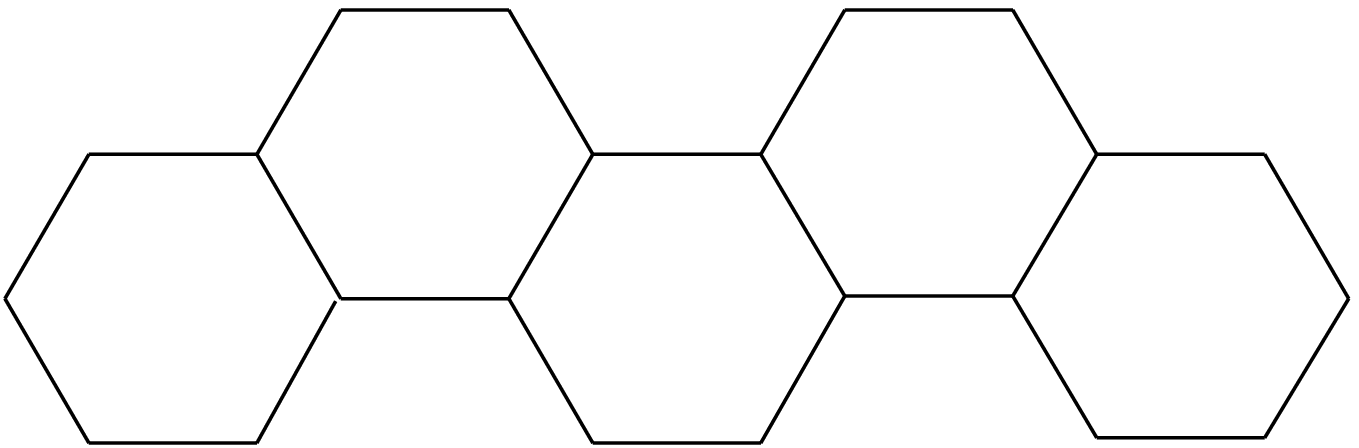}
\includegraphics[height=0.7cm]{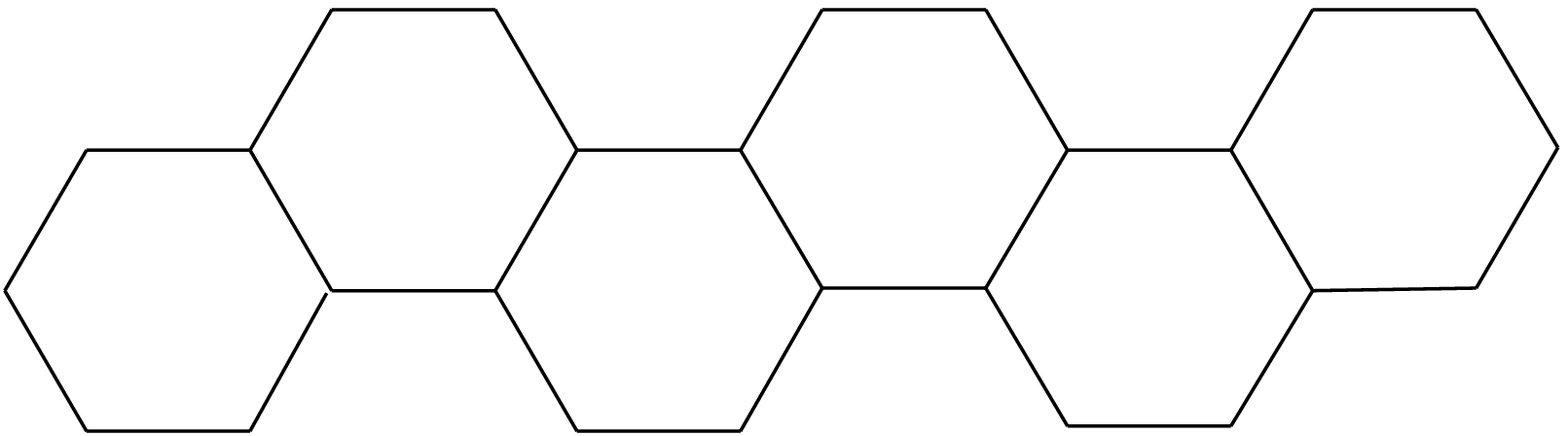}
\includegraphics[height=0.7cm]{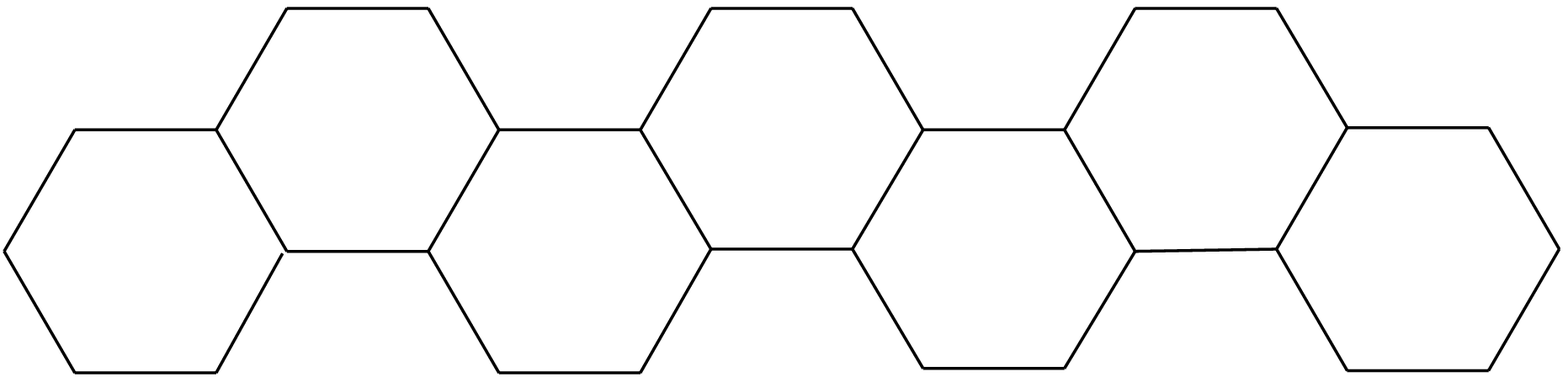}\\[2pt]
\includegraphics[height=1.3cm]{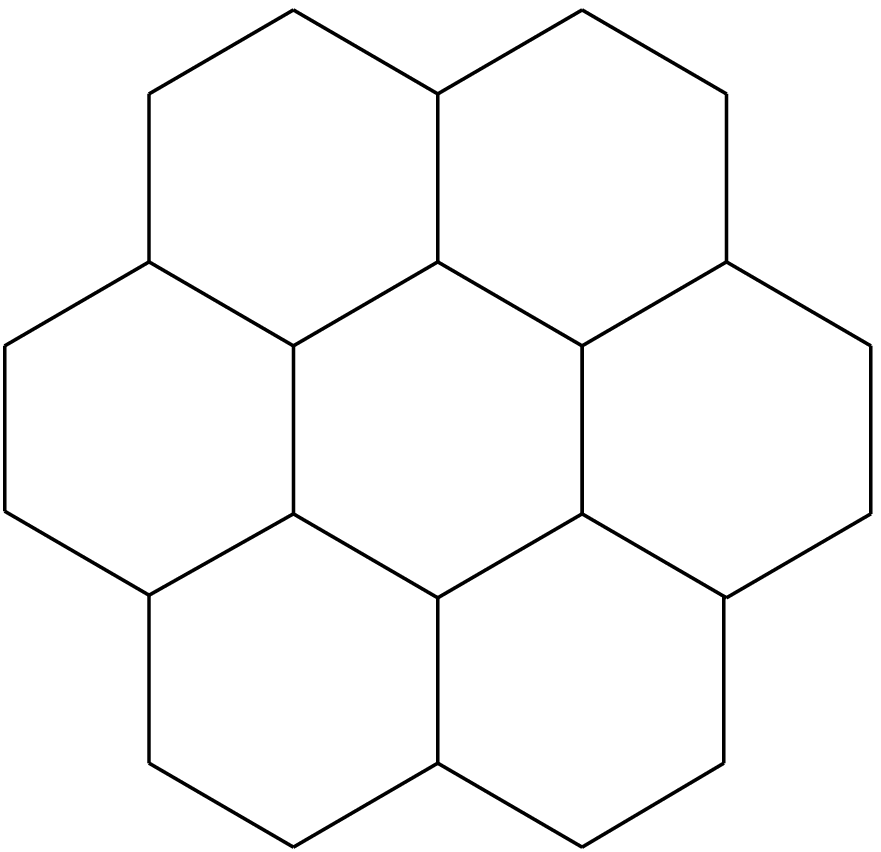}
\includegraphics[height=1.3cm]{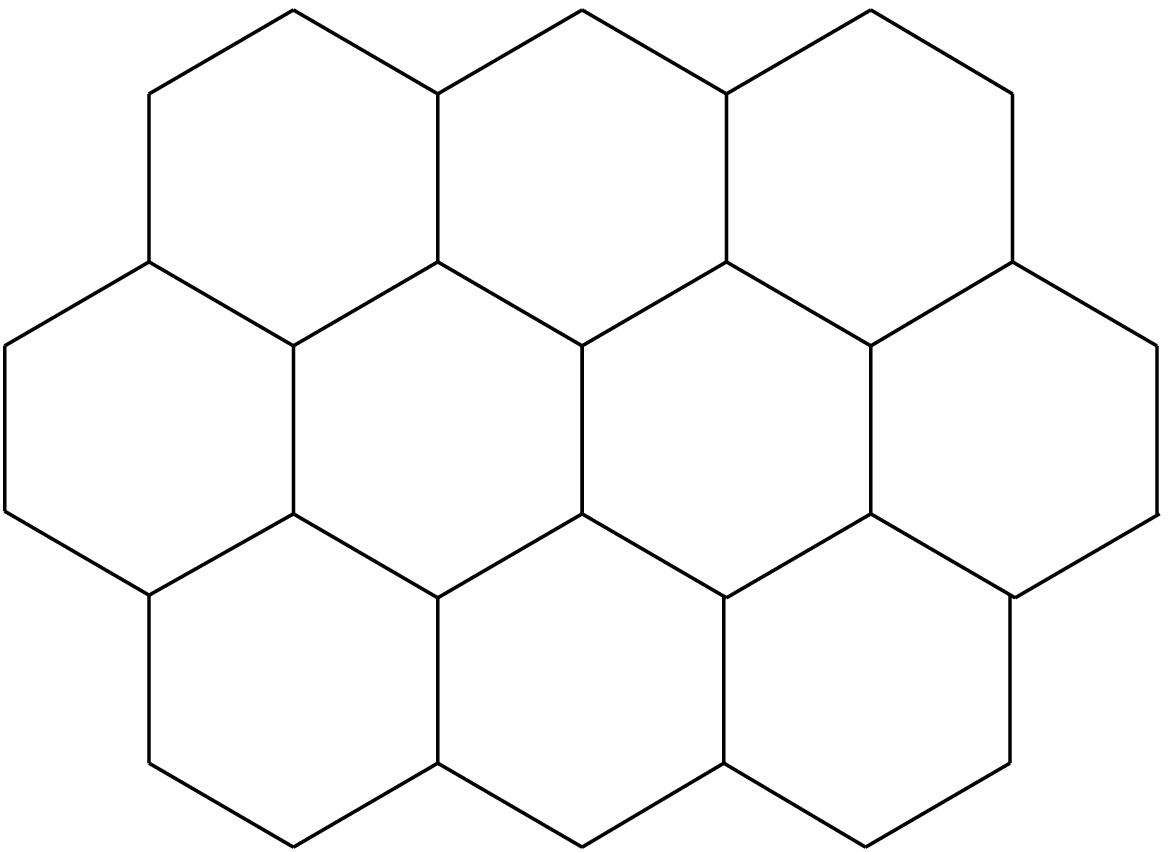}
\includegraphics[height=1.3cm]{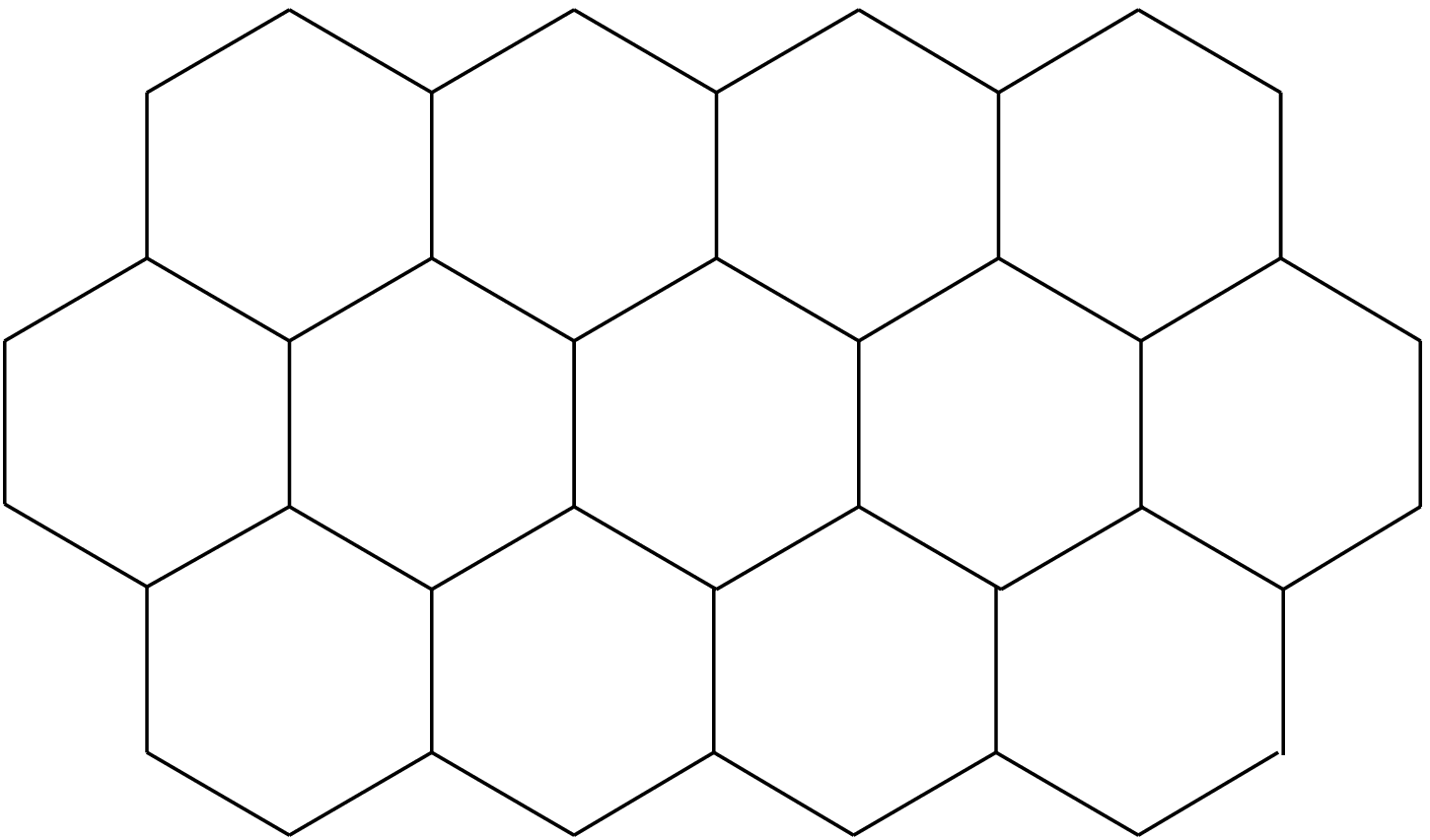}
\includegraphics[height=1.3cm]{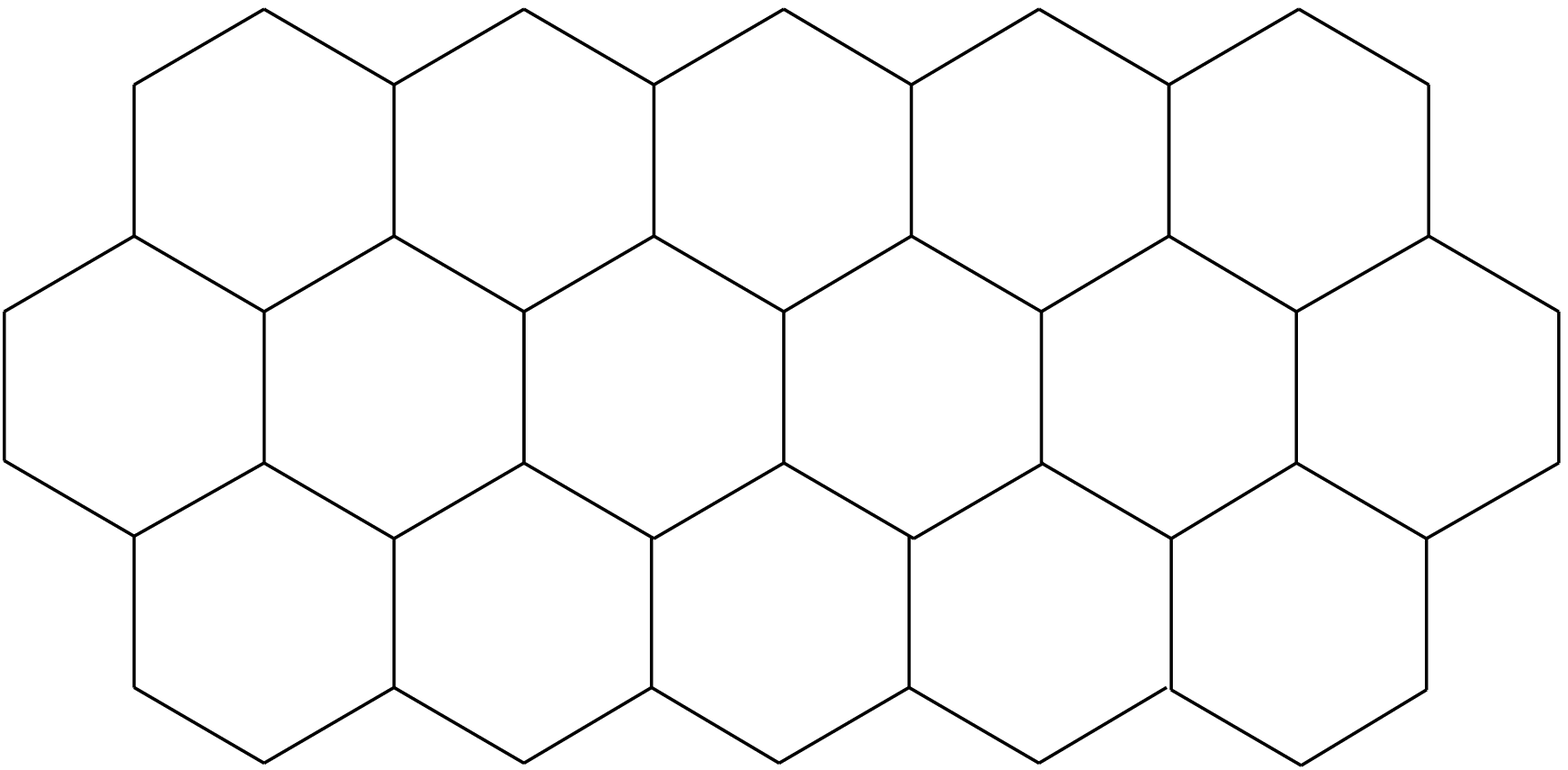}
\includegraphics[height=1.3cm]{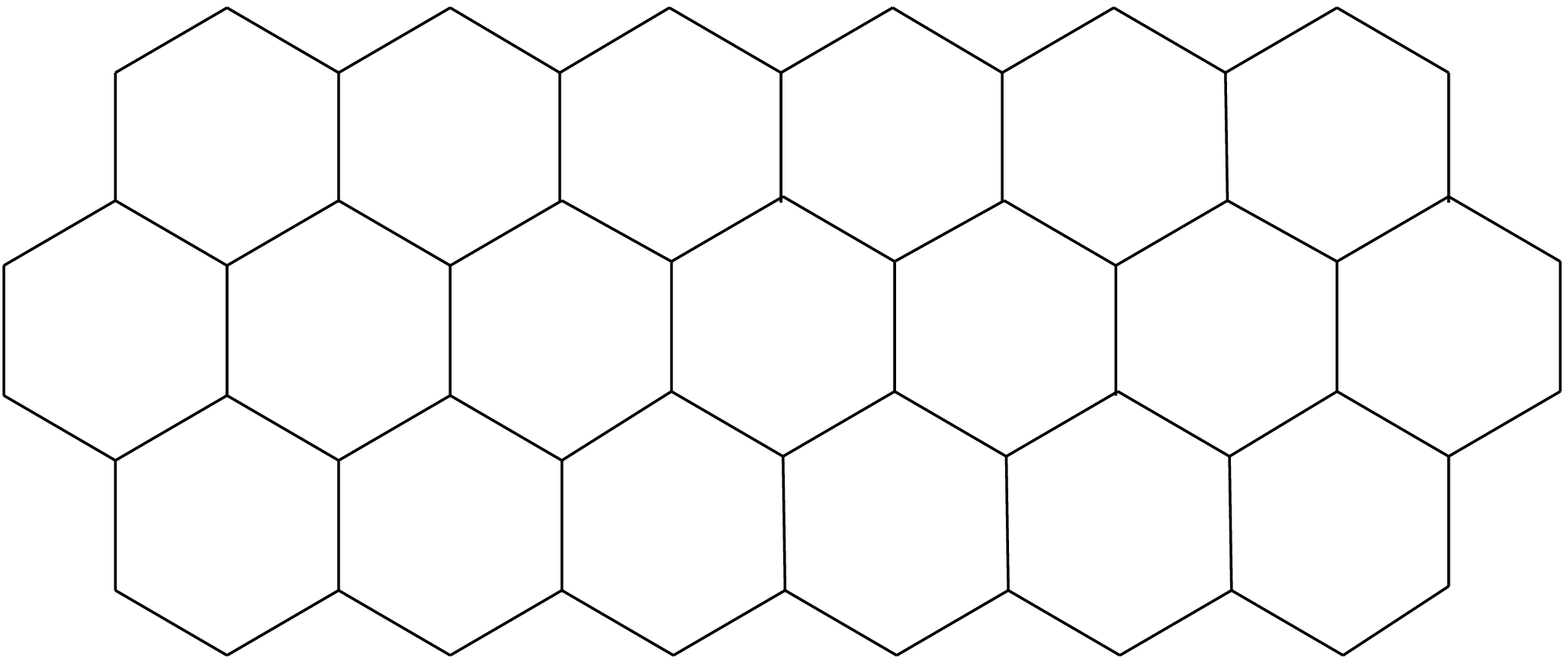}\\[2pt]
\includegraphics[angle=90,height=0.85cm]{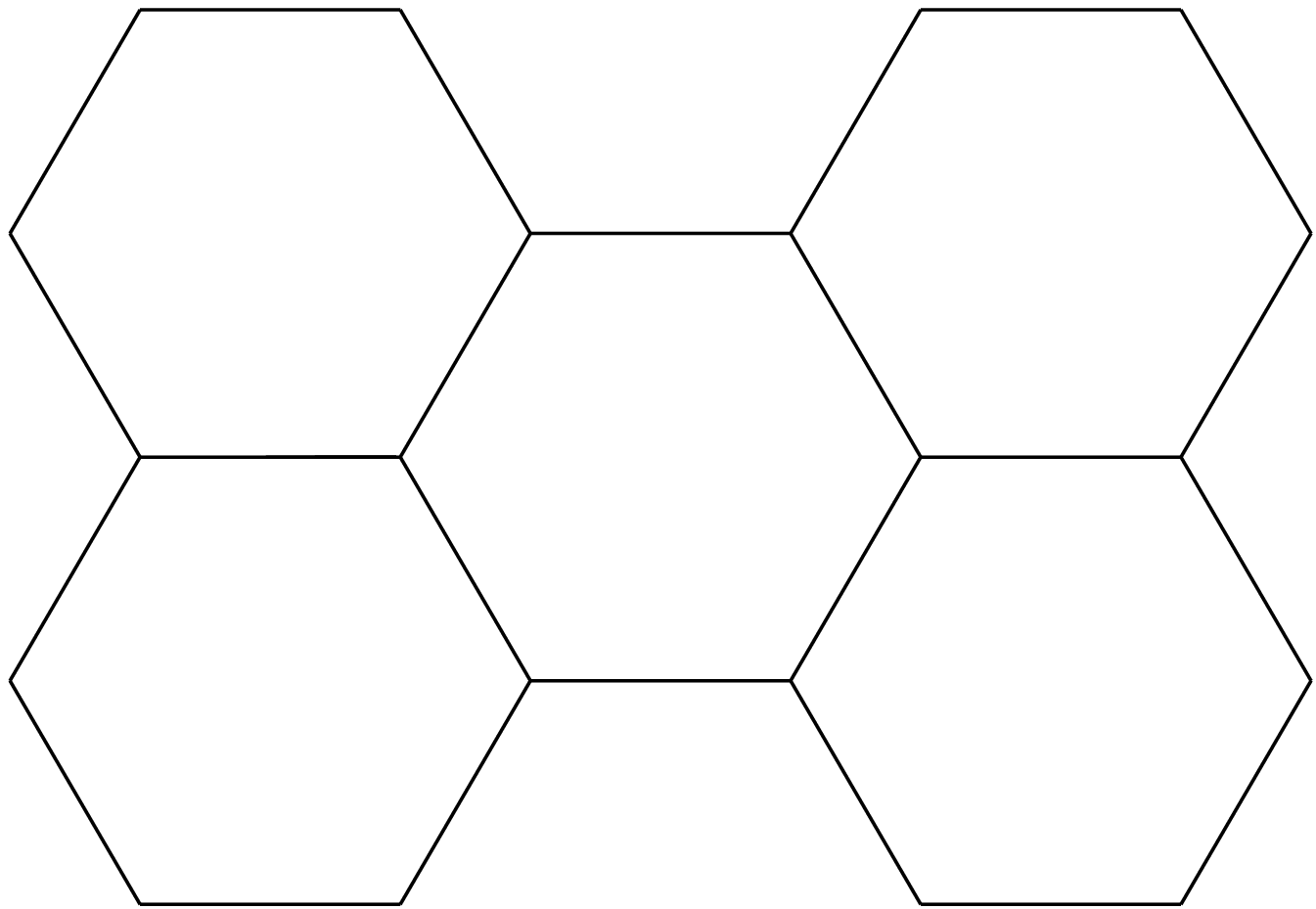}
\includegraphics[angle=90,height=0.85cm]{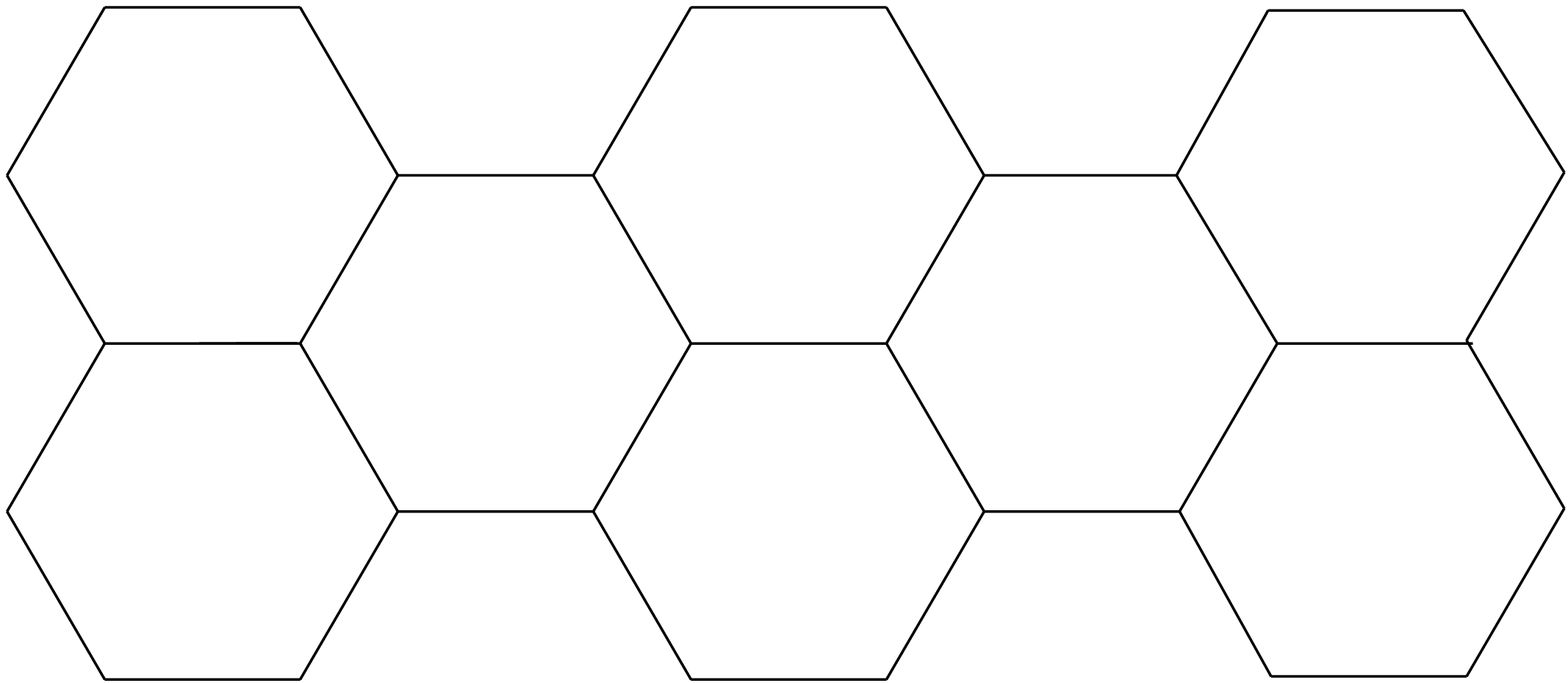}
\includegraphics[angle=90,height=0.85cm]{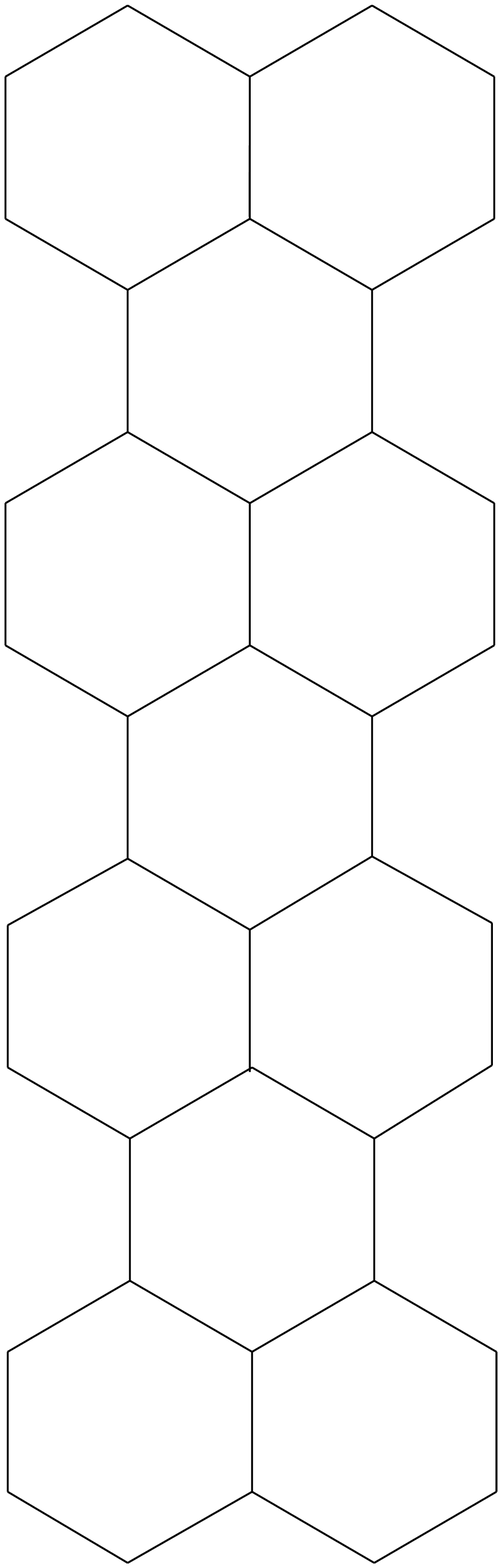}
\includegraphics[angle=90,height=0.85cm]{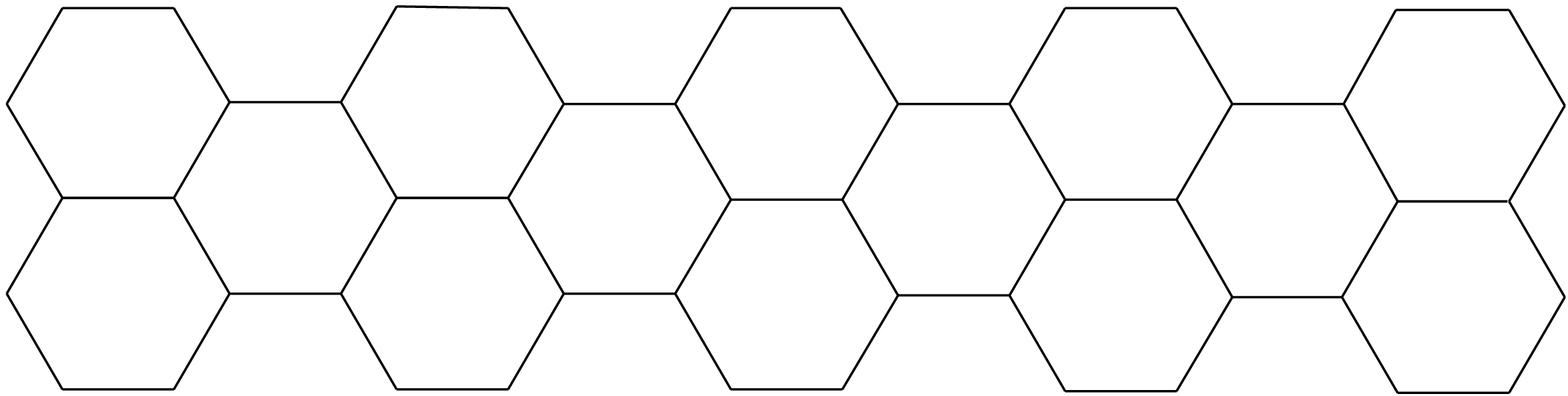}
\includegraphics[angle=90,height=0.85cm]{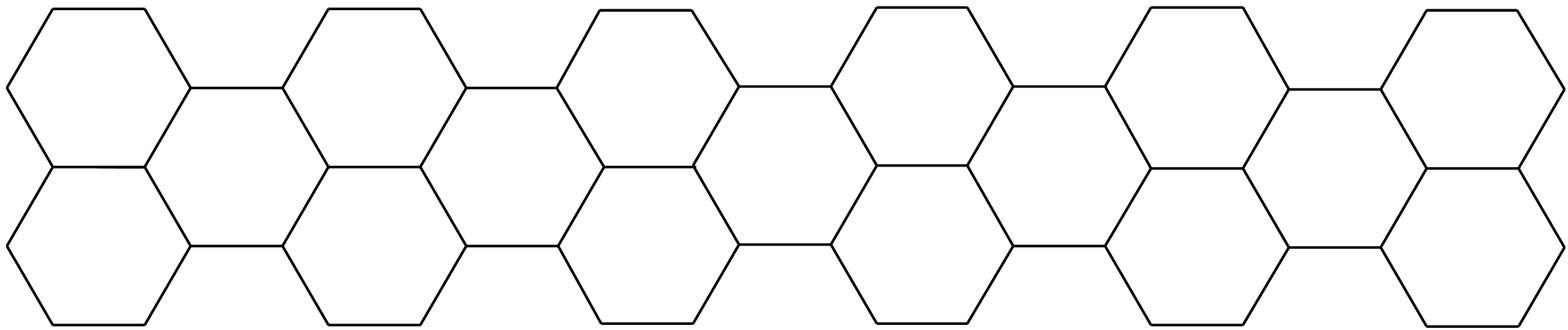}
\caption{Oriented geometries of the molecules considered. 
From top to bottom: oligoacenes (naphthalene, anthracene, tetracene, pentacene, 
hexacene), phenacenes (phenanthrene, chrysene, picene, fulminene, 7-phenacene),
circumacenes (coronene, ovalene, circumanthracene, circumtetracene,
circumpentacene), and oligorylenes (perylene, terrylene, quaterrylene, 
pentarylene, hexarylene). 
\label{molecules}
}
\end{figure}

The geometries of the molecules considered are sketched 
in Fig.~\ref{molecules}. Oligoacenes \cite{ant08} and phenacenes \cite{mal97} 
consist of fused benzene rings joined in a linear and zig\textendash zag 
arrangement, respectively. Oligorylenes \cite{kar98} can be seen as a series of 
naphthalene molecules connected by two carbon\textendash carbon bonds, and 
circumacenes are obtained by addition of benzene units around the circumference 
of their oligoacenes counterparts. Note that, with the exception of 
circumtetracene, circumpentacene, and hexarylene, all of the molecules 
considered have been synthesized.

We used density functional theory (DFT) \cite{god88,jon89,dre90,gor96} to 
compute the electronic
ground\textendash state of the molecules considered, and 
time\textendash dependent DFT (TD-DFT) \cite{hir99,yab99,mar04,dre05} 
to obtain their 
electronic absorption spectra. To the best of our knowledge the complete 
photo\textendash absorption spectra of higher neutral and charged phenacenes, 
circumacenes, and oligorylenes are reported here for the first time. The 
knowledge of this property, in particular, enabled a comparison of the 
absorption in the visible part of the spectrum, a quantity of interest for
photovoltaic applications. Moreover, we used total\textendash energy differences
between the self\textendash consistent field calculations performed for the 
neutral and charged systems ($\pm1$) to evaluate: (i) the vertical and 
adiabatic electron affinities and first ionization energies; (ii) the 
quasiparticle correction to the highest occupied molecular orbital 
(HOMO)\textendash lowest unoccupied molecular orbital (LUMO) gap; (iii) the 
molecular reorganization energies. The comparison between the optical gap, i.e. 
the lowest singlet\textendash singlet TD\textendash DFT excitation energy, and 
the quasiparticle corrected HOMO\textendash LUMO gap enabled a rough estimate 
of the excitonic effects occurring in these molecules.

The paper is organized as follows. Section~\ref{theory} reports some
technical details of the calculations. The results we obtained are presented 
and discussed in Sect.~\ref{results}. Our concluding remarks are reported in 
Sect.~\ref{conclusion}.

\section{Computational details}
\label{theory}
\subsection{ Geometry optimizations}
\label{dft_calculations}

We used the Gaussian\textendash based DFT module of the \textsc{nwchem} package
\cite{nwchem} to compute the ground\textendash state properties of each 
molecule in its $-1, 0, +1$ charge\textendash state. Following our 
previous studies on linear acenes \cite{mal07,cap09}, we used a 
basis\textendash set 
with the smallest addition of diffuse functions, namely the 
\mbox{6\textendash31+G$^\star$} basis, a valence double zeta set augmented with 
$d$ polarization functions and $s$ and $p$ diffuse functions for each carbon 
atom. Geometry optimizations were performed using tight convergence criteria, 
specified by maximum and root mean square gradient thresholds of 
1.5$\cdot$10$^{-5}$ and 1.0$\cdot$10$^{-5}$ atomic units, respectively, 
and maximum and root mean square thresholds of the Cartesian step respectively 
of 6.0$\cdot$10$^{-5}$ and 4.0$\cdot$10$^{-5}$ atomic units. All neutral and 
singly\textendash charged species were computed as singlet and doublet,
respectively, which are known to be the ground states for this class of
molecules \cite{lan96,wib97,rie01}.

{ Geometry optimization were performed using}
the hybrid B3LYP functional, a combination of the Becke's three 
parameter exchange functional \cite{bec93} and the Lee\textendash 
Yang\textendash Parr gradient\textendash corrected correlation functional 
\cite{lee88}. In comparison with other exchange\textendash correlation 
functionals, the B3LYP has proven to give results markedly more accurate for 
the ground\textendash state properties of PAHs \cite{mar96,lan96} and has been 
used in a large number of studies \cite{wib97,des00,rie01,ngu02,par03,gri03,del03,die04,mal05,kat05,kad06}.
{ Although the B3LYP functional has been developed for the electronic  
ground\textendash state, it is also routinely employed to obtain electronic 
excited\textendash state properties, and its application usually yields  
accurate results for low\textendash lying valence\textendash excited states
of both closed\textendash shell \cite{bau96,wib98} and open\textendash shell 
\cite{hir99b,hir99c} species. It is known, however, that this functional show 
the wrong asymptotic behaviour, decaying faster than $1/R$ (i.e., exponentially)
for large distances $R$ from the nuclei. Among the well known and documented 
limitations of conventional functionals \cite{dre05} are: (i) 
the correct description of Rydberg \cite{cas98,toz98,han99}, 
doubly\textendash excited \cite{cav04,mai04}, and 
charge\textendash transfer \cite{toz99,sob03,dre04} 
excited\textendash states, (ii) the failure for large, 
extended $\pi$\textendash systems such as polyacetylene fragments and 
oligoporphyrins \cite{cai02}, and (iii) the 
system\textendash size\textendash dependent errors found for the lowest 
short\textendash polarized excited states of neutral oligoacenes \cite{gri03}.

This latter error in particular, which is particularly relevant to the present 
study, was recently shown to be corrected by range\textendash separated 
functionals \cite{won10}. While conventional hybrid functional such as B3LYP
are defined with a costant fraction of Hartree\textendash Fock exchange, 
the authors of Ref.~\cite{won10} used range\textendash separated functionals 
\cite{iik01,taw04,mun05,sat07} incorporating both a 
position\textendash dependent fraction and an 
asymptotically correct contribution of Hartree\textendash Fock exchange. 
The application of range\textendash separated functionals to organic compounds
show that they provide, in general, a better accuracy than global hybrids
\cite{jac07a,jac07b,jac08,jac09,won09}.}

As to the molecular properties of interest { in this work} (electron 
affinities, ionization energies, quasiparticle corrected HOMO\textendash LUMO 
gaps, exciton binding energy, electronic excitations), we recently checked for 
oligoacenes both the sensitivity of our calculations to different 
exchange\textendash correlation functionals \cite{vos80,pbe}, and their 
reliability in 
comparison with the available experimental data. With the only exception of the 
ionization energy, which were found to be described better at the LDA level, we 
found that the hybrid B3LYP functional yields the overall best performance. 
{ To further asses this choice we performed some benchmark calculations 
using the range separated functional LC\textendash BLYP \cite{won09}, which was 
found to give the best results in describing the optoelectronic properties of 
oligoacenes \cite{won10}. 
Table~\ref{test1} reports the comparison between B3LYP, LC\textendash BLYP, and 
the available experimental data of oligoacenes. A similar comparison for some 
selected properties for which we could find the corresponding experimental data 
is presented in Table~\ref{test2} for the smallest species of the four PAH 
families considered, namely naphthalene, phenanthrene, coronene, and perylene. 

The results in Table~\ref{test1} are consistent with those reported in 
Ref.~\cite{won10}, the small differences being due to the larger basis set 
cc-PVTZ used in the latter work. As shown by the reported mean absolute 
errors (MAE) and root-mean-squared errors (RMS), the use of LC\textendash BLYP
is indeed (i) able to solve the system\textendash size\textendash dependent 
errors of B3LYP found for the lowest excited state corresponding to the 
HOMO\textendash LUMO transition ($p$ bands); (ii) gives ionization energies
that are in much more agreement than B3LYP with experiments; and (iii) provides 
better quasiparticle corrected HOMO-LUMO gaps. At the same time, however, the 
B3LYP functional appears to outperform LC\textendash BLYP with respect to the 
$\alpha$ and $\beta$ bands, the electron affinity, and the exciton binding 
energy; the molecular reorganization energies are described with the same 
accuracy by the two methods. When considering different families as done in 
Table~\ref{test2}, the B3LYP results are always in better agreement with 
experiments, with the only exception of the ionization energy. Since definitive 
conclusions cannot be drawn from our benchmark calculations and a thorough 
comparison between the performances of global hybrids and range\textendash 
separated functionals is beyond the scope of this work,} we decided to extend 
our previous analyses for oligoacenes \cite{mal07,cap09}, using the same level 
of theory B3LYP/\mbox{6\textendash31+G$^\star$} for the other classes of PAHs 
considered in this work. 

\begin{table}
\caption{ Comparison between B3LYP, LC\textendash BLYP, and the available 
experimental data of oligoacenes (number of benzene rings from 2 to 6). The 
three lowest electronic transitions (usually described by Clars's notation $p$, 
$\alpha$, $\beta$) are computed by means of linear response TD\textendash DFT 
(see Sect.~\ref{tddft_calculations}). Ionization energies, electron affinities 
ionization energy, quasi\textendash particle corrected HOMO\textendash LUMO
gaps, exciton binding energy, and molecular reorganization energy are 
computed via total energy differences (see Sect.~\ref{energy_differences}).
The experimental data are taken from the compilation in Ref.~\cite{mal07}
with the exception of the molecular reorganization energies $\lambda$
taken from Ref.~\cite{san07}. The last rows report the mean absolute
errors (MAE) and root-mean-squared errors (RMS) for each quantity; all 
values are given in eV. \label{test1}}
\begin{center}
\begin{tabular}[c]{c c c c c c c c c c}
\hline
\noalign{\smallskip}
n & Method & $p$ & $\alpha$ & $\beta$ & IE & EA & QP & E$_\mathrm{bind}$ & 
$\lambda$ \\
\noalign{\smallskip}
\hline
\noalign{\smallskip}
\multirow{3}*{2} 
& B3LYP & 4.36 & 4.44 & 5.85 & 7.89 & -0.38 & 8.27 & 3.91 & 0.178 \\
& LC-BLYP & 4.82 & 4.60 & 6.07 & 8.13 & -0.42 & 8.55 & 3.95 & 0.217 \\
& EXP & 4.45 & 3.97 & 5.89 & 8.144 & -0.20 & 8.34 & 3.89 & 0.182 \\
\hline
\multirow{3}*{3} 
& B3LYP & 3.21 & 3.85 & 5.14 & 7.09 & 0.43 & 6.66 & 3.45 & 0.134 \\
& LC-BLYP & 3.72 & 4.04 & 5.42 & 7.34 & 0.39 & 6.95 & 3.23 & 0.173 \\
& EXP & 3.45 & 3.84 & 5.24 & 7.439 & 0.53 & 6.91 & 3.46 & 0.174 \\
\hline
\multirow{3}*{4} 
& B3LYP & 2.45 & 3.47 & 4.62 & 6.55 & 1.00 & 5.55 & 3.10 & 0.109\\
& LC-BLYP & 2.98 & 3.67 & 4.95 & 6.80 & 0.96 & 5.85 & 2.87 & 0.150\\
& EXP & 2.72 & 3.12 & 4.55 & 6.970 & 1.067 & 5.90 & 3.18 & 0.138 \\
\hline
\multirow{3}*{5} 
& B3LYP & 1.91 & 3.21 & 4.24 & 6.16 & 1.41 & 4.75 & 2.84 & 0.092\\
& LC-BLYP & 2.44 & 3.42 & 4.60 & 6.42 & 1.37 & 5.05 & 2.61 & 0.133\\
& EXP & 2.31 & 3.73 & 4.40 & 6.589 & 1.392 & 5.20 & 2.89 & 0.102\\
\hline
\multirow{3}*{6} 
& B3LYP & 1.51 & 3.02 & \textemdash & 5.87 & \textemdash & 
\textemdash & \textemdash &  \textemdash \\
& LC-BLYP & 2.00 & 3.20 & \textemdash & 6.13 & \textemdash & 
\textemdash & \textemdash & \textemdash \\
& EXP & 1.90 & 2.80 & \textemdash & 6.360 & \textemdash & 
\textemdash & \textemdash &\textemdash \\
\hline
& \multirow{2}*{B3LYP} & 0.28 & 0.31 & 0.09 & 0.39 & 0.09 & 
0.28 & 0.04 & 0.02\\
MAE & & (0.13) & (0.16)& (0.05)& (0.18)& (0.05) & 
(0.16) & (0.02) & (0.01)\\
(RMS) & \multirow{2}*{LC-BLYP}& 0.23 & 0.42 & 0.24 & 0.14 & 0.12 & 0.11 & 0.22 
& 0.02\\
& & (0.11) & (0.20)& (0.13)& (0.07) & (0.07) & (0.07) & (0.12) & (0.01) \\
\hline
\end{tabular}
\end{center}
\end{table}

\begin{table}
\caption{ Same as in Table~\ref{test1} for the smallest species
of each PAH family considered; the experimental data are taken from
the references given in Tables~\ref{table}-\ref{table1}.
\label{test2}}
\begin{center}
\begin{tabular}[c]{c c c c c c c}
\hline
\noalign{\smallskip}
Molecule & Method & $p$ & IE & EA & QP & E$_\mathrm{bind}$ \\
\noalign{\smallskip}
\hline
\noalign{\smallskip}
\multirow{3}*{Naphthalene} 
& B3LYP & 4.36 & 7.89 & -0.38 & 8.27 & 3.91 \\
& LC-BLYP & 4.82 & 8.13 & -0.42 & 8.55 & 3.95 \\
& EXP & 4.45 & 8.144 & -0.20 & 8.34 & 3.89 \\
\hline
\multirow{3}*{Phenanthrene} 
& B3LYP & 4.19 & 7.63 & -0.21 & 7.84 & 3.65 \\
& LC-BLYP & 4.30 & 7.92 & -0.29 & 8.20 & 3.90 \\
& EXP & 4.24 & 7.891 & 0.01 & 7.88 & 3.64 \\
\hline
\multirow{3}*{Coronene} 
& B3LYP & 3.42 & 7.08 & 0.38 & 6.70 & 3.28 \\
& LC-BLYP & 3.52 & 7.41 & 0.27 & 7.15 & 3.63 \\
& EXP & 3.74 & 7.290 & 0.47 & 6.82 & 3.08 \\
\hline
\multirow{3}*{Perylene} 
& B3LYP & 2.79 & 6.64 & 0.87 & 5.77 & 2.98 \\
& LC-BLYP & 3.26 & 6.92 & 0.81 & 6.11 & 2.85 \\
& EXP & 2.82 & 6.960 & 0.973 & 5.99 & 3.17 \\
\hline
& \multirow{2}*{B3LYP} & 0.12 & 0.26 & 0.15 & 0.11 & 0.10 \\
MAE & & (0.08) & (0.13) & (0.08) & (0.07) & (0.07) \\
(RMS)& \multirow{2}*{LC-BLYP} & 0.27 & 0.05 & 0.22 & 0.25 & 0.30 \\
& & (0.15) & (0.03) & (0.11) & (0.13) & (0.17) \\
\hline
\end{tabular}
\end{center}
\end{table}

\subsection{ Total\textendash energy differences calculations}
\label{energy_differences}

From the structural relaxations performed for both neutral and charged systems,
we computed via total\textendash energy differences the adiabatic electron 
affinities and the adiabatic single ionization energies. At the optimized 
geometry of the neutral molecule we then evaluated the vertical electron 
affinity (EA$_\mathrm{v}$) and the vertical first ionization energy 
(IE$_\mathrm{v}$). This enabled the calculation of the quasiparticle\textendash 
corrected HOMO\textendash LUMO gap of the neutral systems; this quantity is 
usually referred to as the fundamental gap and is rigorously defined within 
the $\Delta$SCF scheme \cite{jon89} as:
\begin{equation}
\label{delta}
\mathrm{QP}_\mathrm{gap}^1 = \mathrm{IE}_\mathrm{v} - \mathrm{EA}_\mathrm{v} = 
\mathrm{E}_\mathrm{N+1} + \mathrm{E}_\mathrm{N-1} - 2 \mathrm{E}_\mathrm{N},
\end{equation}
E$_\mathrm{N}$ being the total energy of the N\textendash electron system. 
We used also the following approximate expression \cite{god88}:
\begin{equation}
\label{delta2}
\mathrm{QP}_\mathrm{gap}^2 = 
\epsilon_\mathrm{N+1}^\mathrm{N+1} - \epsilon_\mathrm{N}^\mathrm{N},
\end{equation}
where $\epsilon_\mathrm{i}^\mathrm{j}$ is the i$^\mathrm{th}$ 
Kohn\textendash Sham eigenvalue of the j\textendash electron system. 
The results obtained using the above 
Eqs.~(\ref{delta}) and (\ref{delta2}) tend to coincide as the system gets 
larger and the orbitals more delocalized.

Concerning transport properties, in organic molecular semiconductors 
conductivity is known to occur via a 
hopping mechanism in which charge carriers jump between adjacent molecules, 
usually under the effect of an external applied field. This process can be 
described by the semiclassical Marcus theory of electron transfer with overall 
rate constant proportional to $\exp{[-\lambda/(4 K_\mathrm{B} T)]}$ 
($K_\mathrm{B}$ Boltzmann's constant, $T$ temperature) \cite{bre04,san07}. The 
quantity $\lambda$, the molecular reorganization energy or intramolecular
coupling, affects critically the charge\textendash transfer process: low
values of the reorganization energy imply high transfer rates. 
$\lambda$ can be computed via the "four\textendash point method" as the sum 
of two contributions 
\cite{bre04,san07}:
\begin{equation}
\label{lambda}
\lambda = \lambda_1 + \lambda_2 = (E^\star_+ - E_+) + (E^\star - E).
\end{equation}
In the equation above, $E$ and $E_+$ are the total energies of the neutral and 
cation in their equilibrium structures, respectively, $E_+^\star$ is the total 
energy of the cation in the neutral geometry, and $E^\star$ is the total energy 
of the neutral in the cation geometry. While the first term $\lambda_1$ 
corresponds to the radical cation formation, the second one $\lambda_2$ 
accounts for the relaxation of the charged state.

\subsection{TD\textendash DFT calculations}
\label{tddft_calculations}

To compute the excitation energies and electronic absorption spectra we used 
two different implementations of TD\textendash DFT in the linear response 
regime, in conjunction with different representations of the wavefunctions:
\begin{enumerate}
\item \label{one}
the frequency\textendash space implementation \cite{hir99} based on the linear
combination of localized orbitals, as given in the \textsc{nwchem}
package \cite{nwchem};
\item \label{two}
the real\textendash time propagation scheme using a grid in real space 
\cite{yab99}, as implemented in the \textsc{octopus} computer program 
\cite{cas02,mar03}.
\end{enumerate}
We used the scheme (\ref{one}) at the same level B3LYP/\mbox{6-31+G$^\star$} 
used to obtain the ground\textendash state geometries and restricted ourselves 
to the first few singlet-singlet electronic transitions. In scheme (\ref{two}) 
one obtains the whole absolute photo\textendash absorption cross\textendash 
section $\sigma(E)$ from the dynamical polarizability $\alpha(E)$ through the 
equation:
\begin{equation}
\label{sigma}
\sigma(E) = \frac{8 \pi^2 E}{h c}\,\Im \{\alpha(E)\},
\end{equation}
where $h$ is Planck's constant, $\Im \{\alpha(E)\}$ is the imaginary part of
the dynamical polarizability, and $c$ is the velocity of light in vacuum. The
dipole strength\textendash function $S(E)$ is related to $\sigma(E)$ by the 
equation:
\begin{equation}
\label{strength}
S(E) = \frac{m_e c} {\pi h e^2} \sigma(E),
\end{equation}
$m_e$ and $e$ being respectively the mass and charge of the electron.
$S(E)$ has units of oscillator strength per unit energy and satisfies
the Thomas\textendash Reiche\textendash Kuhn dipole sum\textendash rule
\mbox{$N_\mathrm{e} = \displaystyle \int\!dE\,S(E)$}, where $N_\mathrm{e}$ is
the total number of electrons. The real\textendash time TD\textendash DFT 
method in real space was proven to give good results for PAHs up to
photon energies of about $30$~eV \cite{mal07,cap09}. For the calculations 
performed 
with \textsc{octopus} we used the same prescription as in Ref. \cite{mal07}. 

\begin{figure}[ht!]
\includegraphics{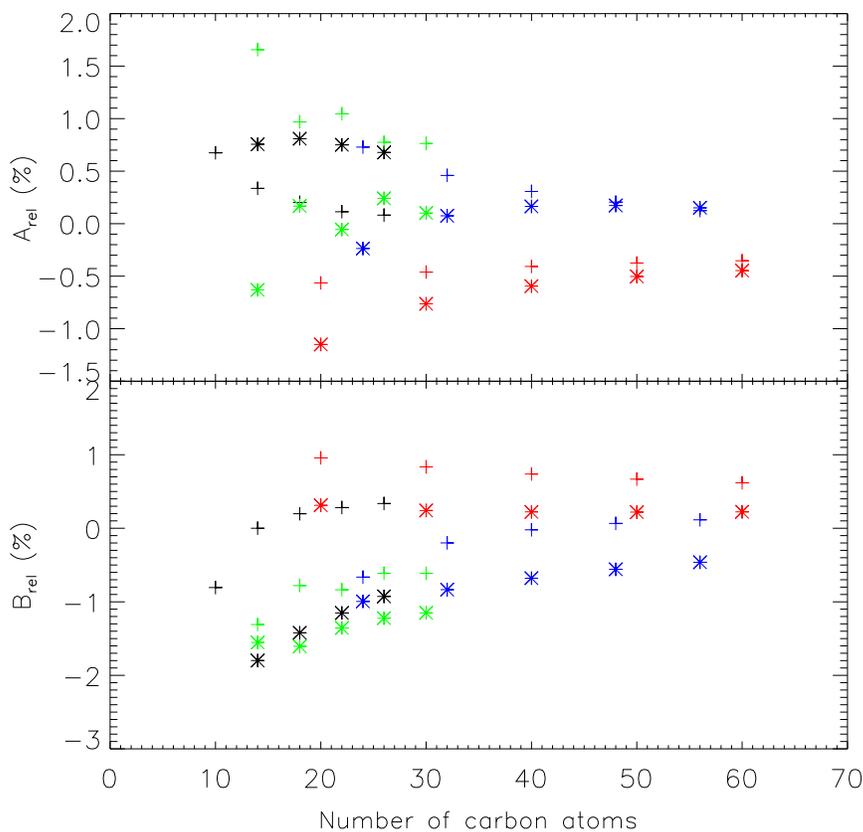}
\caption{Percentage variation of the rotational constants A (top panel)
and B (bottom panel) relative to the neutral counterparts for anions (asterisks)
and cations (crosses), as a function of molecular size for oligoacenes (black), 
phenacenes (green), circumacenes (blue), and oligorylenes (red). We omit the 
entries for naphthalene anion, which is known not to form a stable anion in the 
gas\textendash phase.}
\label{rotconst}
\end{figure}

\section{Results and discussion}
\label{results}

\subsection{Geometry optimization}

In our previous studies \cite{mal07,cap09} we found that the structural parameters 
obtained for ionized oligoacenes, ignoring any apparent symmetry and adopting 
the D$_\mathrm{2h}$ symmetry of the corresponding neutral molecule, are 
coincident within numerical errors. We therefore assumed in this work the 
same symmetry of the neutral molecule for each charged species considered, 
with the exception of coronene anion and cation, which are known to undergo 
Jahn\textendash Teller distortion from D$_{6h}$ to D$_{2h}$ symmetry upon single 
ionization \cite{kat99}. As previously done for oligoacenes \cite{mal07}, the 
structural variations occurring in the charged species compared to the 
corresponding neutral one are expressed in  Fig.~\ref{rotconst} in terms of 
the percentage variations of the rotational constants A and B. These latter 
quantities are proportional to the inverse of the principal momenta of inertia 
in the plane of the molecule.
Since rotational constants are here used
just as a compact way to represent structural changes between different 
charge states, they were all computed at the optimized geometries, disregarding 
vibration\textendash rotation coupling effects.
Evaluating the differences of rotational constants between the charged species 
and the corresponding neutral counterparts, $\Delta$A and $\Delta$B, 
we thus computed the quantities A$_\mathrm{rel}$ = $\Delta$A/A$_\mathrm{neutral}$ 
and B$_\mathrm{rel}$ =  $\Delta$B/B$_\mathrm{neutral}$. Figure~\ref{rotconst} 
shows that A$_\mathrm{rel}$ and B$_\mathrm{rel}$ display well\textendash defined 
trends as a function of molecular size. 

For oligoacenes and circumacenes the 
largest structural changes are observed for the anions, while the variations 
of cations are always small; in particular, the anions of both families appear 
to be primarily distorted along the long axis showing for all sizes a 
lengthening along it (B$_\mathrm{rel}<0$) in comparison to the neutral molecules.
The same holds true for both $\pm1$ charge\textendash states of phenacenes, 
the cations showing also a sensible shortening along the short\textendash axis 
(A$_\mathrm{rel}>0$). In the case of oligorylenes, instead, at increasing 
molecular size the largest structural changes are observed for the cations, 
which appear to be primarily distorted along the long axis showing a shortening 
along it (B$_\mathrm{rel}>0$).

\begin{table}
\caption{Adiabatic and vertical values (within parentheses) of electron 
affinities and ionization energies of oligoacenes, phenacenes, circumacenes, 
and oligorylenes; all data, in eV, have been obtained via total energy 
differences at the \mbox{B3LYP/6\textendash31+G$^\star$} level. 
The experimental data are taken from the NIST Chemistry WebBook \cite{lia05}.
\label{table}}

\begin{center}
\begin{tabular}{ccccc}
\hline \hline
\multirow{2}*{Molecule} & \multicolumn{2}{c}{Electron affinity} & 
\multicolumn{2}{c}{Ionization energy}\\
 & Theory & Experiment & Theory & Experiment\\
\hline
\multicolumn{5}{c}{Acenes (C$_{4n+2}$H$_{2n+4}$)} \\ 
naphthalene & -0.26(-0.38) & -0.20$\pm$0.05 & 7.80(7.89) & 8.144$\pm$0.001 \\
anthracene & 0.53(0.43) & 0.530$\pm$0.005 & 7.02(7.09) & 7.439$\pm$ 0.006 \\
tetracene & 1.08(1.00) & 1.067$\pm$0.043 & 6.49(6.55) & 6.97$\pm$0.05 \\
pentacene & 1.48(1.41) & 1.392$\pm$0.043 & 6.12(6.16) & 6.589$\pm$0.001 \\
hexacene & 1.78(1.72) & \textemdash{} & 5.83(5.87) & 6.36$\pm$0.02\\
\hline
\multicolumn{5}{c}{Phenacenes (C$_{4n+2}$H$_{2n+4}$)} \\
phenanthrene & -0.05(-0.21) & 0.01$\pm$0.04$^a$ & 7.53(7.63) & 
7.891$\pm$0.001 \\
chrysene & 0.29(0.19) & 0.397$\pm$0.008 & 7.17(7.25) & 7.60$\pm$0.01\\
picene & 0.40(0.10) & 0.542$\pm$0.008 & 7.04(7.13) & 7.51$\pm$0.02\\
fulminene & 0.57(0.48) & \textemdash{} & 6.88(6.95) & 7.36$\pm$0.02\\
7\textendash phenacene & 0.64(0.54) & \textemdash{} & 6.80(6.87) & 
\textemdash{} \\
\hline
\multicolumn{5}{c}{Circumacenes (C$_{8n}$H$_{2n+6}$)} \\
coronene & 0.47(0.38) & 0.470$\pm$0.090 & 7.02(7.08) & 7.29$\pm$0.03\\
ovalene & 1.17(1.11) & \textemdash{} & 6.36(6.41) & 6.71$\pm$0.01\\
circumanthracene & 1.68(1.63) & \textemdash{} & 5.90(5.94) & \textemdash\\
circumtetracene & 2.06(2.02) & \textemdash{} & 5.57(5.60) & \textemdash{} \\
circumpentacene & 2.34(2.31) & \textemdash{} & 5.33(5.35) & \textemdash{} \\
\hline
\multicolumn{5}{c}{Rylenes (C$_{10n}$H$_{4n+4}$)} \\
perylene & 0.96(0.87) & 0.973$\pm$0.005 & 6.57(6.64) & 6.960$\pm$0.001\\
terrylene & 1.55(1.48) & \textemdash{} & 5.98(6.05) & 6.42$\pm$0.02\\
quaterrylene & 1.91(1.85) & \textemdash{} & 5.62(5.68) & 6.11$\pm$0.02 \\
pentarylene & 2.17(2.11) & \textemdash{} & 5.37(5.43) & \textemdash{} \\
hexarylene & 2.37(2.31) & \textemdash{} & 5.18(5.24) & \textemdash\\
\hline
\end{tabular}
\end{center}
$^a$ Taken from Ref.~\cite{tsc06}.
\end{table}

\begin{figure}[b!]
\includegraphics{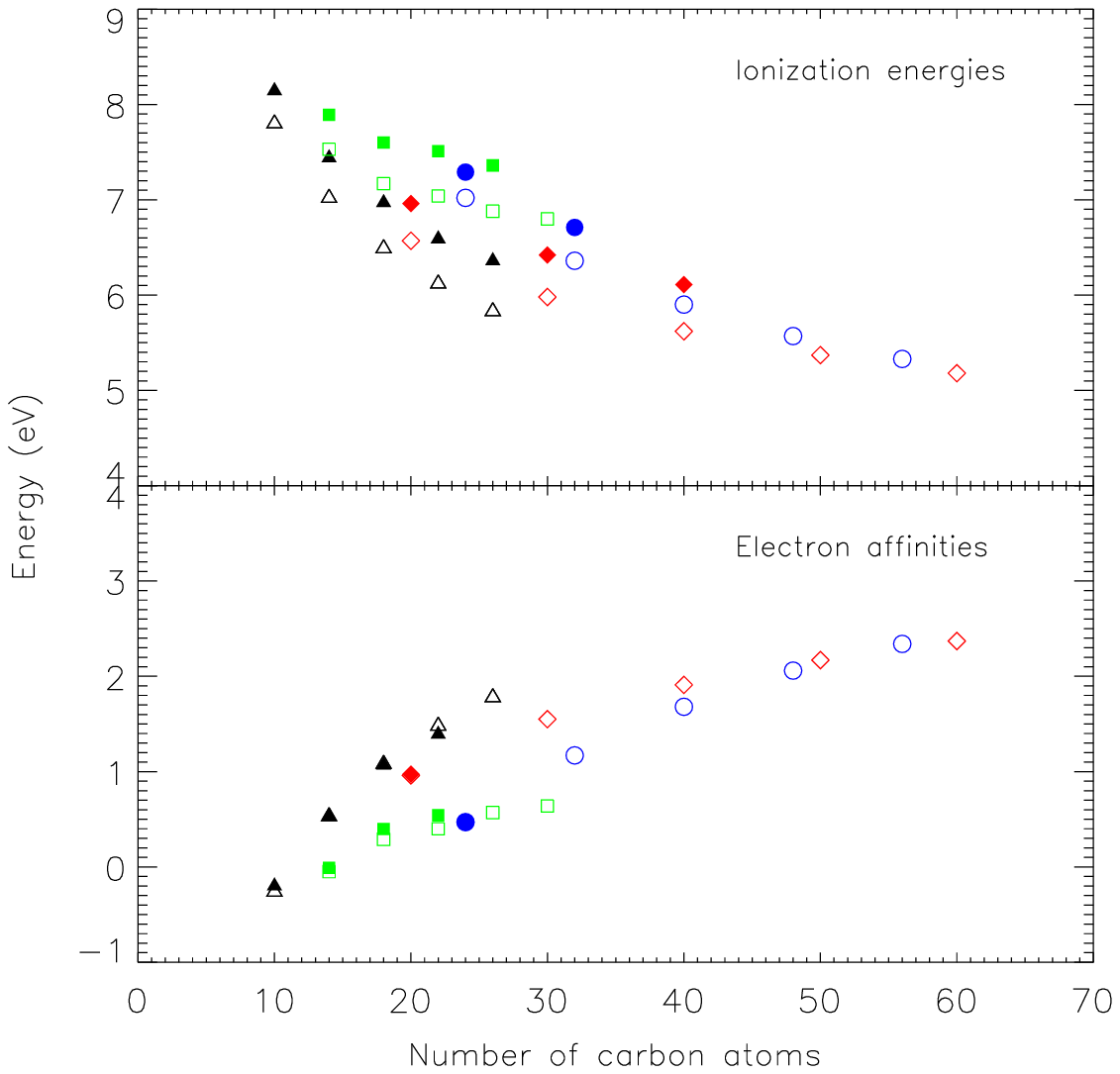}
\caption{Computed adiabatic ionization energies and electron affinities as 
a function of size for oligoacenes (black triangles), phenacenes 
(green squares), circumacenes (blue circles), and oligorylenes (red diamonds). 
When available, the corresponding experimental values are represented by the 
filled symbols (cfr.~Table~\ref{table}).}
\label{ioniz}
\end{figure}

\subsection{Ionization energies and electron affinities}
\label{ionization}

The adiabatic and vertical values of electron affinities and ionization 
energies as obtained via total energy differences at the 
\mbox{B3LYP/6-31+G$^\star$} level are given in Table~\ref{table} and displayed 
in Fig.~\ref{ioniz} as a function of molecular size; for comparison, we report 
also the available experimental data \cite{lia05}. In particular, the 
differences between computed and measured first ionization energies, 
underestimated by theory in all cases, are of the same order of magnitude 
as in previous analyses \cite{wib97,kad06}: they increase systematically at 
increasing molecular size for each class (from $\sim$4\% for naphthalene to 
$\sim$10\% for hexacene, from $\sim$5\% for phenanthrene to $\sim$7\% for 
fulminene, from $\sim$6\% for perylene to $\sim$8\% for quaterrylene, and 
from $\sim$4\% for coronene to $\sim$5\% for ovalene). { As already
found in the literature \cite{won10} and confirmed in our benchmark 
calculations (see Sect.~\ref{dft_calculations}, 
Tables~\ref{test1}-\ref{test2}), these errors in the 
ionization energies can be largely corrected by range\textendash separated 
functionals. On the contrary,} as compared to ionization energies, the 
computed electron affinities are found to be in better agreement with 
experiments in all cases, confirming the very good performances of 
\mbox{B3LYP/6\textendash31+G$^\star$} to compute the electron affinities of 
PAHs \cite{des00,rie01,mal05}. Oligoacenes display the largest variations for 
ionization energies and electron affinities as a function of molecular size. 
On the other hand, phenacenes show the smallest variations for the same 
properties. Circumacenes and oligorylenes behave in the same way at increasing 
sizes, with ionization energies reaching the value of about 5.2~eV and electron 
affinities converging to about 2.4~eV.

\begin{figure}
\includegraphics{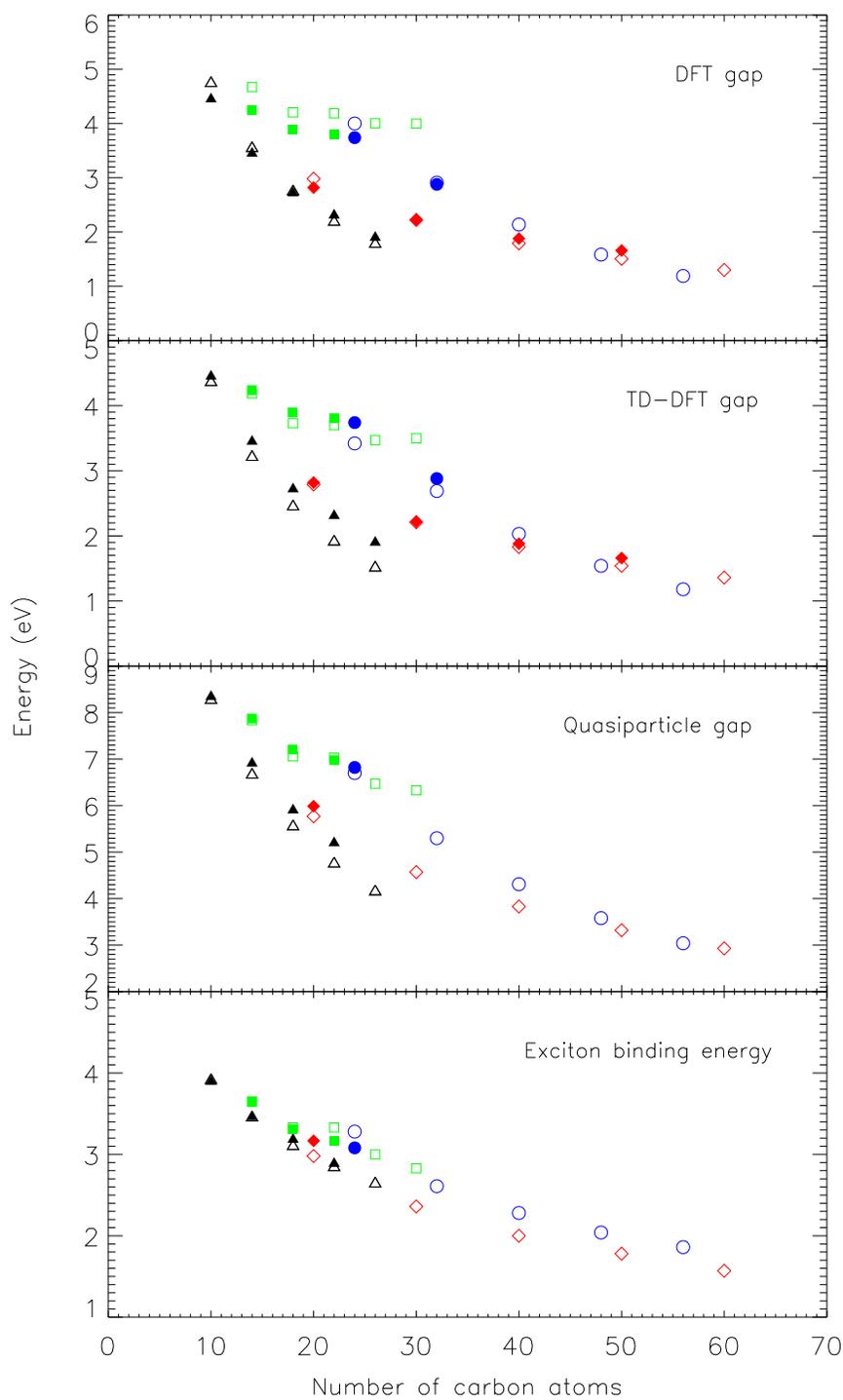}
\caption{Computed HOMO\textendash LUMO, TD\textendash DFT, and quasiparticle 
corrected energy gaps, and exciton binding energy as a function of molecular 
size for oligoacenes (black triangles), phenacenes (green squares), 
circumacenes (blue circles), and oligorylenes (red diamonds). When available, 
the corresponding experimental values are represented by the filled symbols 
(cfr.~Table~\ref{table1}).}
\label{trends}
\end{figure}

\subsection{HOMO\textendash LUMO gap and excitonic effects}
\label{homo-lumo}

For each molecule considered, Table~\ref{table1} compares the 
HOMO\textendash LUMO gap E$_\mathrm{gap}^\mathrm{KS}$ obtained as difference of 
Kohn-Sham eigenvalues, the excitation energy of the HOMO\textendash LUMO 
transition E$_\mathrm{gap}^\mathrm{TD-DFT}$ as given by frequency\textendash 
space TD\textendash DFT, and the corresponding experimental value 
E$_\mathrm{gap}^\mathrm{exp}$. In Table~\ref{table1} we compare also the 
quasiparticle\textendash corrected HOMO\textendash LUMO gap computed
via Eqs.~(\ref{delta}) and (\ref{delta2}) to the corresponding 
experimental value QP$_\mathrm{gap}^\mathrm{exp}$ = 
IE$_\mathrm{exp}$-EA$_\mathrm{exp}$ (cfr.~Table~\ref{table}). 
The theoretical exciton binding energy E$_\mathrm{bind}$ is estimated through 
the difference QP$^1_\mathrm{gap}$-E$_\mathrm{gap}^\mathrm{TD-DFT}$, and 
compared with its corresponding experimental value 
QP$_\mathrm{gap}^\mathrm{exp}$ - E$_\mathrm{gap}^\mathrm{exp}$.
For each class considered all of these quantities are displayed in 
Fig.~\ref{trends} and, as expected due to a reduction of quantum confinement 
effects, they all decrease as a function of molecular size.

As shown in Table~\ref{table1} and Fig.~\ref{trends} in the case of oligoacenes 
the HOMO\textendash LUMO gap obtained as difference between 
Kohn\textendash Sham eigenvalues, with relative errors in the range 1-6\%, 
gives a better description of the experimental optical gap as compared to 
TD\textendash DFT. In the latter case, { as already mentioned before},  
a system\textendash size\textendash dependent error is known to exist 
\cite{gri03}, with a relative error increasing from 2 to 20\% when going from 
naphthalene to hexacene. For circumacenes and oligorylenes the two methods 
have approximately the same accuracy in predicting the optical gap, with 
relative errors in the range 1-8\%. On the contrary, TD\textendash DFT results 
for phenacenes agree better with experiment (relative error 1-4\%) than the 
corresponding Kohn\textendash Sham gap (relative error 8-10\%).

As previously discussed in Ref.~\cite{mal07}, the $\Delta$SCF 
QP\textendash corrected HOMO\textendash LUMO gaps of neutral oligoacenes 
compare very well with the DFT\textendash based tight\textendash binding 
GW data \cite{nie05} but they are affected by similar errors as those found 
with TD\textendash DFT. On the other hand, since these errors almost cancel 
each other in the evaluation of the exciton binding energy, we obtain an 
accurate estimate of E$_\mathrm{bind}$. To the best of our knowledge the 
QP\textendash corrected HOMO\textendash LUMO gap for phenacenes, circumacenes, 
and oligorylenes have never been reported before in the literature. By 
comparing these values with the corresponding optical gap predicted by 
TD\textendash DFT we could estimate the excitonic effects occurring in these 
molecules. We found in all cases appreciable excitonic effects due to both 
quantum confinement and reduction of screening; E$_\mathrm{bind}$ decreases as 
the size of the molecule increases with similar slope for the different 
families and approaches the value of about 1.6\textendash 1.8 eV for the 
largest species considered.

\begin{figure}
\includegraphics{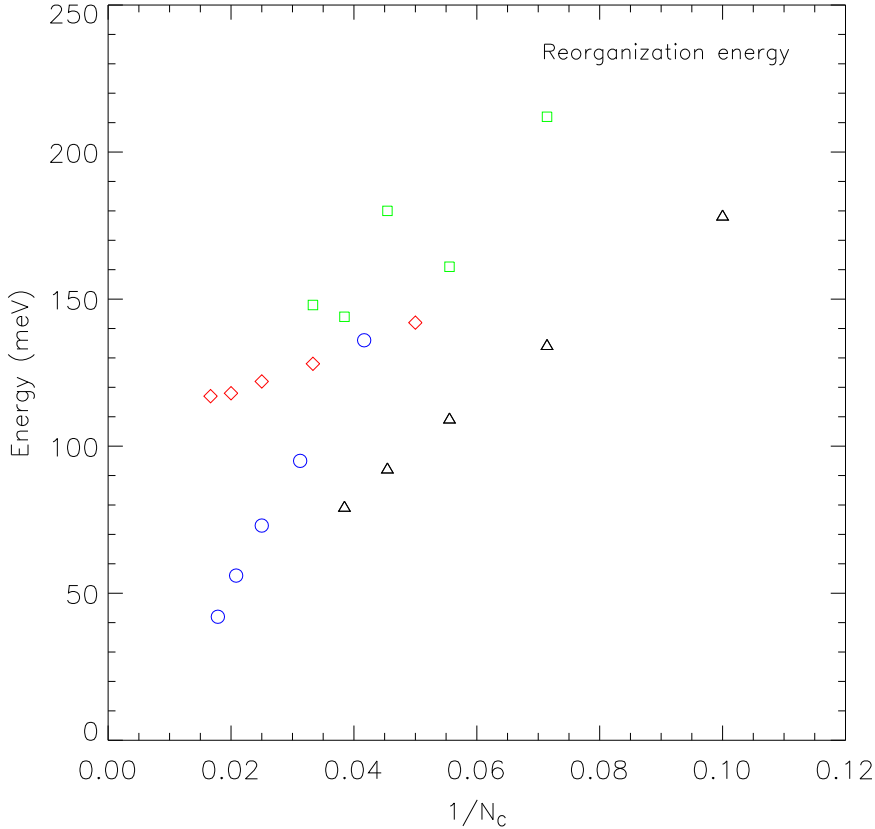}
\caption{Molecular reorganization energies computed using Eq.~\ref{lambda} for 
oligoacenes (black triangles), phenacenes (green squares), circumacenes 
(blue circles), and oligorylenes (red diamonds), as a function of the
inverse of  the total number of carbon atoms $\mathrm{N}_\mathrm{C}$.} 
\label{reorg}
\end{figure}

\subsection{Molecular reorganization energies}
\label{reorganization}

The last row of Table~\ref{table1} reports for each molecule the
molecular reorganization energies $\lambda$ obtained through Eq.~\ref{lambda}. 
Our results for oligoacenes are in good agreement with previous DFT results
\cite{bre04,san07}. Note that, while the performances of the B3LYP functional 
for 
the calculation of $\lambda$ in conjugated materials can be improved by 
precisely calibrated DFT functionals \cite{san07}, its use in the present work 
is justified since it is expected to describe reasonably the differences 
between different families of PAHs. Due to strong electron\textendash phonon 
coupling, the molecular reorganization energy $\lambda$ is known to be inversely
proportional to the total number of carbon atoms $\mathrm{N}_\mathrm{C}$ 
\cite{dev98}.  Figure~\ref{reorg} confirms this general trend with the only
exception of phenacenes, which display for $\lambda$ an oscillatory behaviour 
like for the other physical properties computed (see 
Figs.~\ref{rotconst}-\ref{trends}). For the different families we observe a 
markedly different behaviour: while the reorganization 
energy of oligorylenes { decreases} slowly as a function of molecular size, 
circumacenes display a much more pronounced slope as compared to oligoacenes. 
These results are consistent with those previously found in Ref.~\cite{san09}, 
and confirm that circumacenes are indeed good candidates for new emerging 
technologies.

\begin{landscape}
\begin{table}
\caption{Comparison between HOMO\textendash LUMO gap, 
TD\textendash DFT excitation energy of the HOMO\textendash LUMO transition, 
quasi\textendash particle corrected gap, and molecular reorganization energy 
for the PAHs considered. All data have been obtained at the 
B3LYP/\mbox{6\textendash31+G$^\star$} level of theory.}
\begin{center}
\begin{tabular}{c c c c c c c c c c}
\hline \hline
Molecule (formula) & E$_\mathrm{gap}^\mathrm{KS}$ 
& E$_\mathrm{gap}^\mathrm{TD-DFT}$ & 
E$_\mathrm{gap}^\mathrm{exp}$ & 
QP$^1_\mathrm{gap}$ & QP$^2_\mathrm{gap}$ & 
QP$^\mathrm{exp}_\mathrm{gap}$ & 
E$_\mathrm{bind}$ & E$_\mathrm{bind}^\mathrm{exp}$
& $\lambda$ \\[2pt]
\hline 
\hline
naphthalene (C$_{10}$H$_{8}$) & 4.74 & 4.36 & 4.45$^a$ & 8.27 & 8.12 & 8.34 & 
3.91 & 3.89 & 0.178 \\
anthracene (C$_{14}$H$_{10}$) & 3.54 & 3.21 & 3.45$^a$ & 6.66 & 6.58 & 6.91 & 
3.45 & 3.46 & 0.134 \\
tetracene (C$_{18}$H$_{12}$) & 2.74 & 2.45 & 2.72$^a$ & 5.55 & 5.50 & 5.90 & 
3.10 & 3.18 & 0.109 \\ 
pentacene (C$_{22}$H$_{14}$) & 2.19 & 1.91 & 2.31$^a$ & 4.75 & 4.72 & 5.20 & 
2.84 & 2.89 & 0.092 \\
hexacene (C$_{26}$H$_{16}$) & 1.78 & 1.51 & 1.90$^a$ & 4.15 & 4.13 & 
\textemdash{} & 2.64 & \textemdash{} & 0.079 \\
\hline
phenanthrene (C$_{14}$H$_{10}$) & 4.67 & 4.19 & 4.24$^b$ & 7.84 & 7.44 & 7.88 & 
3.65 & 3.64 & 0.212 \\
chrysene (C$_{18}$H$_{12}$) & 4.21 & 3.73 & 3.89$^b$ & 7.06 & 7.00 & 7.20 & 
3.33 & 3.31 & 0.161 \\
picene (C$_{22}$H$_{14}$) & 4.19 & 3.70 & 3.80$^b$ & 7.03 & 6.82 & 6.97 & 
3.33 & 3.17 & 0.180 \\
fulminene (C$_{26}$H$_{16}$) & 4.00 & 3.47 & \textemdash{} & 6.47 & 6.42 & 
\textemdash{} & 3.00 & \textemdash{} & 0.144 \\
7\textendash phenacene (C$_{30}$H$_{18}$) & 4.00 & 3.50 & \textemdash{} & 6.33 & 
6.26 & \textemdash{} & 2.83 & \textemdash{} & 0.148 \\
\hline
coronene (C$_{24}$H$_{12}$) & 4.00 & 3.42 & 3.74$^c$ & 6.70 & 6.64 & 6.82 & 
3.28 & 3.08 & 0.136 \\
ovalene (C$_{32}$H$_{14}$) & 2.91 & 2.69 & 2.88$^c$ & 5.30 & 5.28 & 
\textemdash{} & 2.61 & \textemdash{} & 0.095 \\
circumanthracene (C$_{40}$H$_{16}$) & 2.14 & 2.03 & \textemdash{} & 4.31 & 4.30 &
\textemdash{} & 2.28 & \textemdash{} & 0.073 \\
circumtetracene (C$_{48}$H$_{18}$) & 1.58 & 1.54 & \textemdash{} & 3.58 & 3.58 & 
\textemdash{} & 2.04 & \textemdash{} & 0.056 \\
circumpentacene (C$_{56}$H$_{20}$) & 1.19 & 1.18 & \textemdash{} & 3.04 & 3.04 & 
\textemdash{} & 1.86 & \textemdash{} & 0.042 \\
\hline
perylene (C$_{20}$H$_{12}$) & 2.98 & 2.79 & 2.82$^d$ & 5.77 & 5.71 & 5.99 & 
2.98 & 3.17 & 0.142 \\
terrylene (C$_{30}$H$_{16}$) & 2.22 & 2.21 & 2.22$^d$ & 4.57 & 4.53 & 
\textemdash{} & 2.36 &\textemdash& 0.128 \\
quaterrylene (C$_{40}$H$_{20}$) & 1.79 & 1.83 & 1.88$^d$ & 3.83 & 3.81 & 
\textemdash{} & 2.00 &\textemdash& 0.122 \\
pentarylene (C$_{50}$H$_{24}$) & 1.51 & 1.54 & 1.66$^d$ & 3.32 & 3.30 & 
\textemdash{} & 1.78 & \textemdash{} & 0.118 \\
hexarylene (C$_{60}$H$_{28}$) & 1.30 & 1.36 & \textemdash& 2.93 & 2.92 & 
\textemdash{} & 1.57 & \textemdash{} & 0.117 \\
\hline
\end{tabular}
\end{center}
$^a$ See compilation in Ref.~\cite{mal07}. 
$^b$ From Ref.~\cite{par03}. 
$^c$ From Ref.~\cite{job92}. 
$^d$ From Ref.~\cite{kar98}.
\label{table1}
\end{table}
\end{landscape}

\subsection{Absorption spectra}

\begin{figure}
\includegraphics{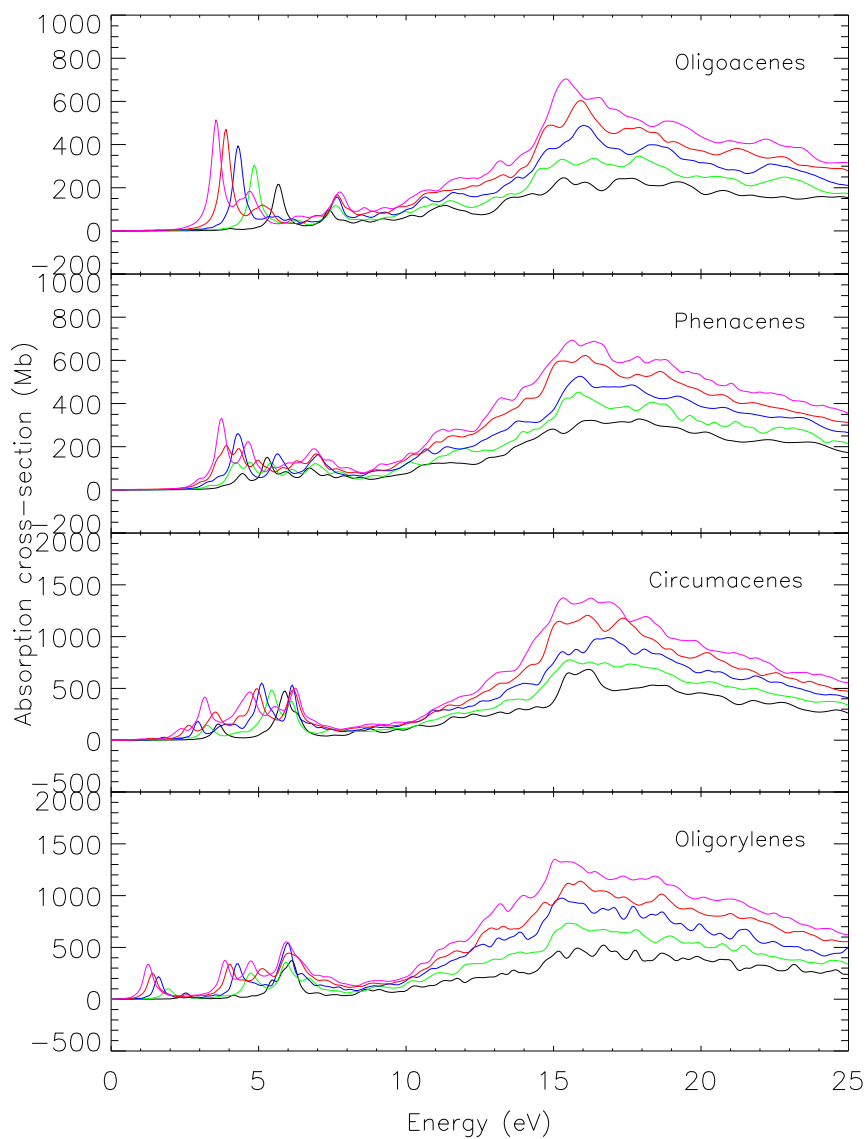}
\caption{Electronic absorption spectra of the neutral molecules of the
four PAH classes considered, as computed using the real\textendash time 
real\textendash space TD\textendash DFT implementation of the 
\textsc{octopus} code.}
\label{spectra}
\end{figure}

\begin{figure}
\includegraphics{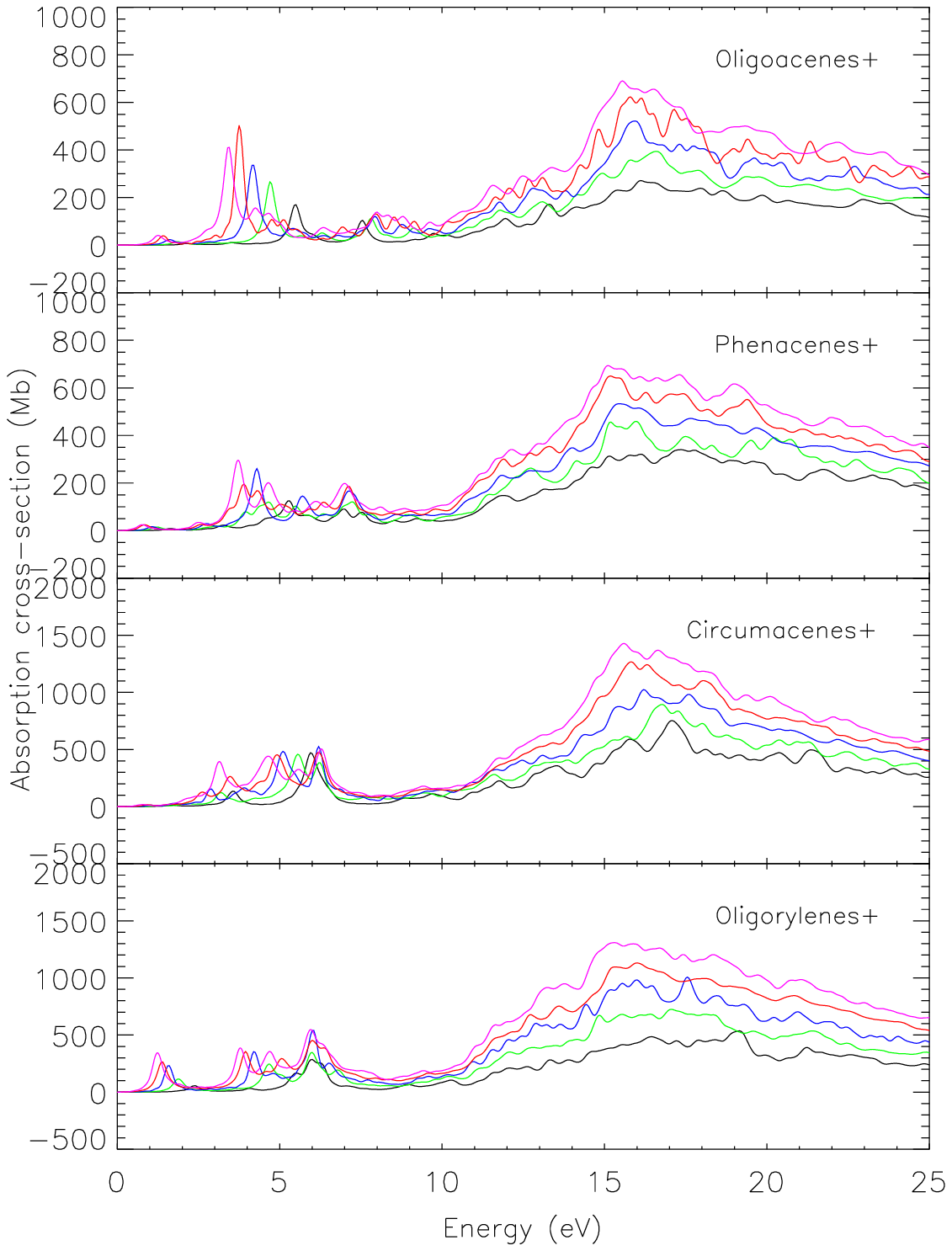}
\caption{Same as Fig.~\ref{spectra} for the corresponding cations.}
\label{spectra+}
\end{figure}

\begin{figure}
\includegraphics{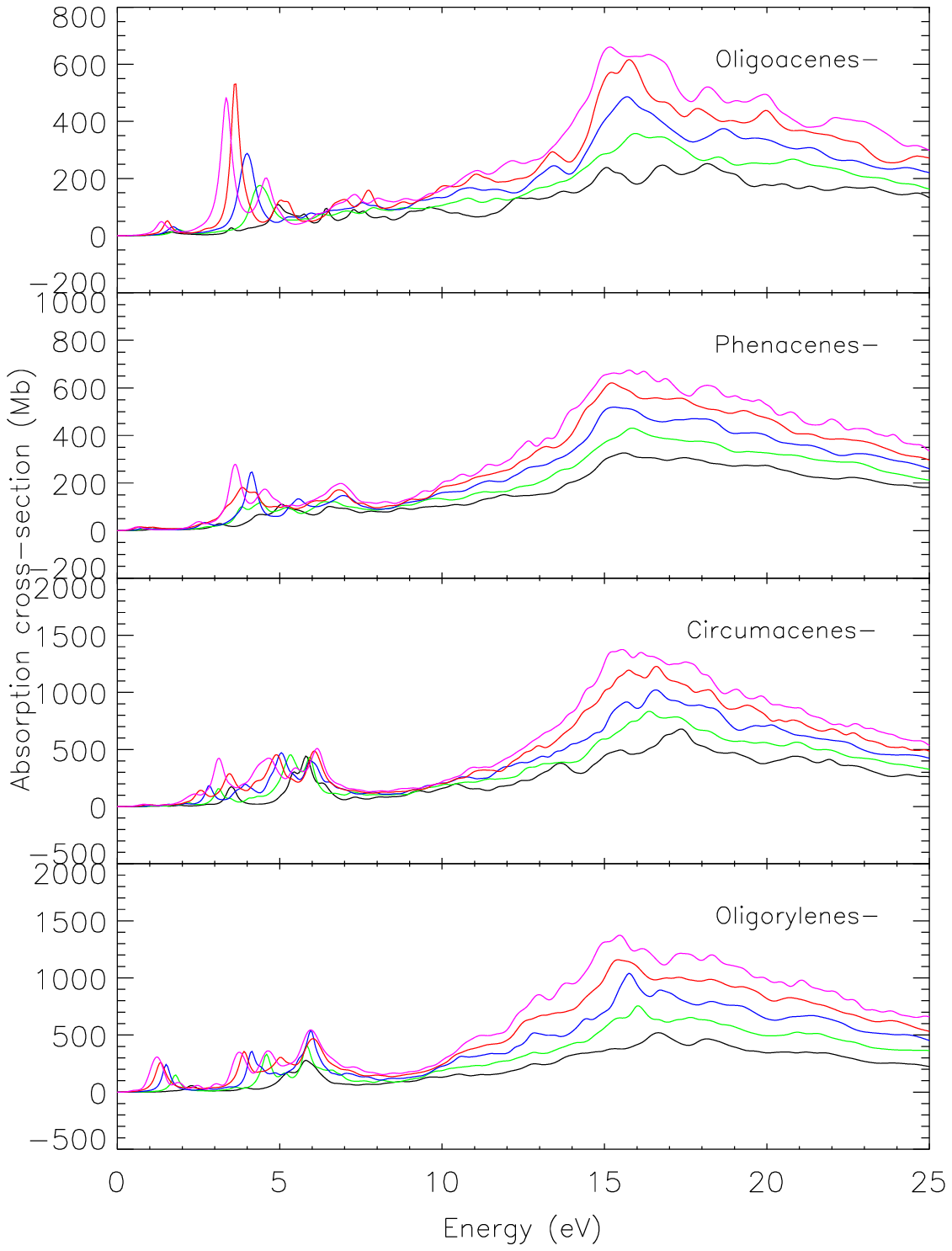}
\caption{Same as Fig.~\ref{spectra} for the corresponding anions.}
\label{spectra-}
\end{figure}

\begin{figure}
\includegraphics{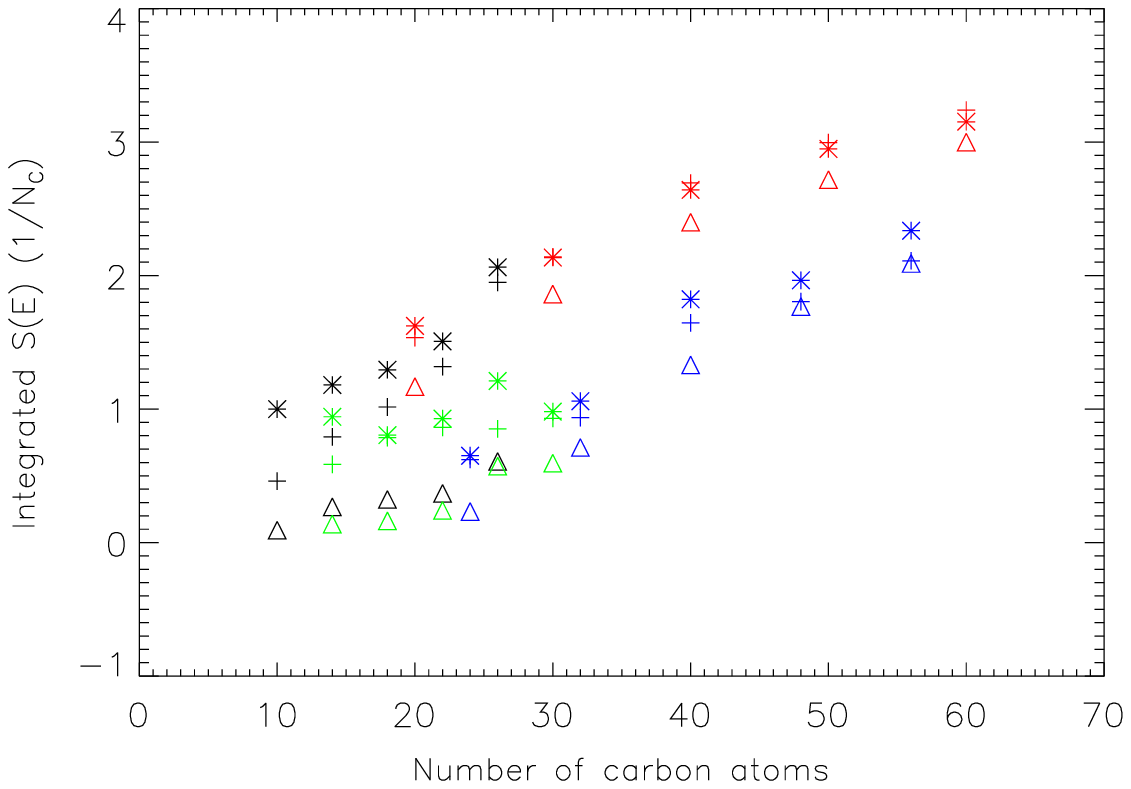}
\caption{Comparison between the integrated values in the range 
1.0\textendash3.0~eV of the individual dipole strength\textendash 
functions $S(E)$ (see Eq.~\ref{strength})divided by the total number of 
carbon atoms N$_\mathrm{C}$ for the anions (asterisks), neutrals (triangles), 
cations (crosses) of oligoacenes (black), phenacenes (green), circumacenes 
(blue), and oligorylenes (red) as a function of molecular size.}
\label{integrated}
\end{figure}

Figures~\ref{spectra}-\ref{spectra-} display the absorption spectra of each 
of the molecule considered, in its neutral and $\pm1$ charge-state, as 
computed using the real\textendash time real\textendash space 
TD\textendash DFT implementation \cite{cas02,mar03}. Note that the spectra of 
the 
higher neutral and singly\textendash charged phenacenes, circumacenes, and 
oligorylenes, have never been reported before. This enabled a quantitative 
comparison of the absorption in the visible part of the spectrum for each 
class, which gives an estimate of its potential use as photo\textendash active 
element in opto\textendash electronic devices. To this end Fig.~\ref{integrated}
shows the integrated values in the range 1.0\textendash3.0~eV of the individual 
dipole strength\textendash functions $S(E)$ (see Eq.~\ref{strength}) divided by 
the total number of carbon atoms in each molecule as a function of molecular 
size. 

As expected from previous analyses \cite{mal07,cap09}, with the exception of 
the different intensities which are related to the total number of electrons 
in the molecule, all of the spectra in Figs.~\ref{spectra}-\ref{spectra-} 
display a 
similar broad excitation peaking at $\sim$17\textendash18~eV, which involves 
$\pi\to\sigma^\star$, $\sigma\to\pi^\star$, $\sigma\to\sigma^\star$, and Rydberg 
spectral transitions, and which is relatively insensitive to the 
charge\textendash state of the molecule. Note that we already showed 
\cite{mal07,cap09} 
our results to be in good agreement up to photon energies of about $30$~eV with 
the experimental data obtained for a few neutral PAHs \cite{job92}. 
However, the use of a finite simulation box in our simulations does not give 
a satisfactory treatment of continuum effects in this spectral window and 
produces spurious structures as compared to experiments \cite{mal07,cap09}. 

The main differences among families arise in the low\textendash energy part of 
the spectrum, from the near\textendash IR to the near\textendash UV up to about 
8~eV, which involves electronic transitions of $\pi\to\pi^\star$ character. 
This is also the spectral window which is more interesting from an applicative
point of view. In particular, an { inspection} of Fig.~\ref{spectra} shows 
that each family displays a continuous redshift of the absorption at increasing 
sizes. This behaviour was expected as a consequence of the continuous
decrease in bandgap already observed for each class (see Fig.~\ref{trends}). 
As to the charged species, as shown in 
Figs.~\ref{spectra+}\textendash \ref{spectra-} we found that PAH radical 
cations and anions display intense optical transitions at lower energies than 
their parent molecule, and have very similar electronic spectra. These 
findings, already known for charged PAHs from extensive spectroscopic studies 
in frozen glassy organic solids \cite{shi73,shi88}, are in qualitative agreement with 
the particle\textendash hole equivalence in the pairing theorem of H\"uckel's 
theory. 

The behaviour described above at low energies translates for 
each molecule in each class into a systematic increase of the integrated 
strength\textendash function in the range 1.0\textendash3.0~eV when going from 
the neutral to the cation and anion (see Fig.~\ref{integrated}). This effect is 
more pronounced for oligoacenes and phenacenes and, as expected, the observed 
scatter between the different charge\textendash states decreases with increasing
molecular size.  Interestingly, oligorylenes appear to be much more efficient 
in absorbing low\textendash energy photons in comparison to the other classes;
as an example, focusing on the neutral species with the same number of carbon 
atoms, the value of S(E)/N$_\mathrm{C}$ for quaterrylene exceeds by more than
80\% the corresponding value for circumanthracene. Since red photons represent 
a significant portion of the solar energy, this result might have important 
technological implications.

\section{Concluding remarks}
\label{conclusion}

We presented a systematic theoretical study of the five smallest species of 
oligoacenes, phenacenes, circumacenes, and oligorylenes in their -1, 0, and +1
charge\textendash states. From the ground\textendash state structural 
relaxations performed at the \mbox{B3LYP/6-31+G$^\star$} level we computed the 
electron affinities, the first ionization energies, the molecular reorganization
{ energies}, and the quasiparticle correction to the HOMO\textendash LUMO 
energy gap. We found good agreement with the available experimental data as 
well as with previous theoretical results. { Benchmark calculations 
performed using the range\textendash separated functional LC\textendash BLYP
suggest that the agreement with the experiment can be further improved for 
some selected properties (e.~g. the ionization energy); a systematic study
will be the subject of future work}. Concerning transport properties, 
circumacenes appear to have the most favorable properties for technological 
applications, displaying a steeper decrease of the molecular reorganization 
energy as the molecular size increases. We computed also the electronic 
absorption spectra using a compendium of the TD\textendash DFT theoretical 
framework in both frequency space, to obtain the optical gap and thus estimate 
the exciton binding energy, and real\textendash time real\textendash space, to 
obtain in a single step the whole photo\textendash absorption 
cross\textendash section extending up to the far\textendash UV. The results of 
this latter method, in particular, enabled a close comparison between the 
families considered with respect to their potential use as photoactive 
materials in opto\textendash electronic devices. Overall, oligorylenes appear 
to be much more efficient in absorbing visible light and red photons in 
particular, which might have important technological implications in the field 
of photovoltaics.

\section*{Acknowledgements}
This work has been funded by the Italian Institute of Technology (IIT) 
under Project SEED ``POLYPHEMO''. We acknowledge computational support
by CYBERSAR (Cagliari, Italy) and CASPUR (Rome, Italy). 
{ G.~C. acknowledges financial support from IDEA\textendash AISBL.}

\end{document}